\documentclass[letterpaper,titlepage,11pt]{article}

\pdfoutput=1

\usepackage{amssymb,amsmath,amsfonts}
\usepackage{mathrsfs}
\usepackage{epsfig}
\usepackage{ccaption}
\usepackage{graphicx}
\usepackage{todonotes}
\usepackage{upgreek}

\usepackage[
      colorlinks=true,
      linkcolor=blue,
      urlcolor=magenta,
      filecolor=green,
      citecolor=red,
      pdfstartview=FitV,
      pdftitle={},
        pdfauthor={Daniel Grumiller, Jakob Salzer, Dmitry Vassilevich},
        pdfsubject={},
        pdfkeywords={},
        pdfpagemode=None,
        bookmarksopen=true
      ]{hyperref}

\numberwithin{equation}{section}

\def\clock{{\count0=\time
           \divide\count0 60
           \ifnum\count0<10 0\fi\the\count0
           \multiply\count0 -60 \advance\count0 \time
           :\ifnum\count0<10 0\fi \the\count0
         }}
\newcommand{\timestamp}{{\small\vbox{\hbox{\tt\jobname.tex}
\hbox{\the\day/\the\month/\the\year, \clock}}}}

\usepackage{amsmath,amssymb}
\usepackage{verbatim}
\usepackage{graphicx}
\usepackage{hyperref}
\usepackage{color}

\DeclareFontFamily{OT1}{rsfs}{}
\DeclareFontShape{OT1}{rsfs}{m}{n}{ <-7> rsfs5 <7-10> rsfs7 <10->rsfs10}{} 
\DeclareMathAlphabet{\mycal}{OT1}{rsfs}{m}{n}

\renewcommand{\dag}{\dagger}

\newcommand{\be}[1]{ \begin{equation}\label{#1} }
\newcommand{\ee}{\end{equation}}
\newcommand{\bea}[1]{\begin{eqnarray}\label{#1} }
\newcommand{\eea}{\end{eqnarray}}

\newcommand{\eq}[2]{\begin{equation} #1 \label{#2} \end{equation}}

\newcommand{\ga}{\gamma}
\newcommand{\de}{\delta}
\newcommand{\om}{\omega}

\newcommand{\la}{\lambda}

\DeclareMathOperator{\extdm}{d}
\newcommand{\extd}{\extdm \!}
\newcommand{\vp}{\varphi}

\newcommand{\del}{\mathrm{d}}

\newcommand{\mat}{\theta} % THE MATRIX FORMALLY KNOWN AS \beta

\begin{document}

%\pacs{...}

%\maketitle

\begin{titlepage}
%\leftline{}{\timestamp}
\leftline{}{TUW--15--20}
\vskip 3cm
\centerline{\LARGE \bf AdS$\boldsymbol{_2}$ holography is (non-)trivial for (non-)constant dilaton} 
%\medskip
%\centerline{\LARGE \bf }
\vskip 1.6cm
\centerline{\bf Daniel Grumiller$^a$, Jakob Salzer$^a$, and Dmitri Vassilevich$^{b}$}
\vskip 0.5cm
\centerline{\sl $^a$Institute for Theoretical Physics, TU Wien}
\centerline{\sl Wiedner Hauptstrasse 8-10/136, A-1040 Vienna, Austria}
\vskip 0.5cm
\centerline{\sl $^b$CMCC-Universidade Federal do ABC, Santo Andr\'e, S.P., Brazil}
\vskip 0.5cm
\centerline{\small\tt \href{mailto:grumil@hep.itp.tuwien.ac.at}{grumil@hep.itp.tuwien.ac.at}, \href{mailto:salzer@hep.itp.tuwien.ac.at}{salzer@hep.itp.tuwien.ac.at}, \href{mailto:dvassil@gmail.com}{dvassil@gmail.com}}

\vskip 1.6cm
\centerline{\bf Abstract} \vskip 0.2cm \noindent

We study generic two-dimensional dilaton gravity with a Maxwell field and prove its triviality for constant dilaton boundary conditions, despite of the appearance of a Virasoro algebra with non-zero central charge. We do this by calculating the canonical boundary charges, which turn out to be trivial, and by calculating the quantum gravity partition function, which turns out to be unity. We show that none of the following modifications changes our conclusions: looser boundary conditions, non-linear interactions of the Maxwell field with the dilaton, inclusion of higher spin fields, inclusion of generic gauge fields. Finally, we consider specifically the charged Jackiw--Teitelboim model, whose holographic study was pioneered by Hartman and Strominger, and show that it is non-trivial for certain linear dilaton boundary conditions. We calculate the entropy from the Euclidean path integral, using Wald's method and exploiting the chiral Cardy formula. The macroscopic and microscopic results for entropy agree with each other.

\end{titlepage}
\pagestyle{empty}
\small
%\tableofcontents
\normalsize
\newpage
\pagestyle{plain}
\setcounter{page}{1}

%\listoftodos

\addtocontents{toc}{\protect\setcounter{tocdepth}{3}}

\tableofcontents
\newpage

\section{Introduction}\label{se:1}

Two is the lowest dimension where a lightcone exists. Therefore, the simplest toy models for classical and quantum gravity that can have black hole solutions and a non-trivial causal structure are two-dimensional. Einstein gravity, however, is not among them: in two dimensions the Einstein tensor vanishes identically for any metric, the Einstein--Hilbert action is a boundary term and the formal counting of physical degrees of freedom yields a meaningless $-1$.

There are various ways to see that dilaton gravity is the model of choice in two dimensions, see the list after Eq.~(15) in \cite{Grumiller:2014qma}. In the present work we study generic dilaton gravity with a Maxwell field in two dimensions, with the (Euclidean) bulk action 
\eq{
I = -\frac{k}{4\pi}\,\int\extd^2x\sqrt{g}\,\big(XR - U(X)(\partial X)^2 - 2V(X) - \tfrac14\,F(X)f_{\mu\nu}f^{\mu\nu}\big)
}{eq:intro0}
where $k=1/(4G_N)$ is inversely proportional to the Newton constant, $U$, $V$, $F$ are arbitrary functions of the dilaton field $X$, and $f_{\mu\nu}=\partial_\mu a_\nu-\partial_\nu a_\mu$ is the abelian field strength, see \cite{Strominger:1994tn, Grumiller:2002nm, Grumiller:2006rc} for reviews and \cite{Strobl:1994eu, Louis-Martinez:1995rq} for some early literature on dilaton gravity with gauge fields. 

The best-known examples are the Jackiw--Teitelboim model \cite{Jackiw:1984, Teitelboim:1984} ($U=F=0$, $V\propto X$), the Witten black hole \cite{Mandal:1991tz, Elitzur:1991cb, Witten:1991yr, Klebanov:1991qa, Dijkgraaf:1992ba} ($U=-1/X$, $V\propto X$, $F=0$), the CGHS model \cite{Callan:1992rs} ($U$, $V$, $F$ either like Witten black hole or $U=F=0$, $V=\textrm{const.}$, plus scalar matter), Liouville gravity \cite{Ginsparg:1993is, Nakayama:2004vk} ($U=\textrm{const.}$, $V\propto e^{\alpha X}$, $F=0$), type 0A/0B string theory \cite{Douglas:2003up, Gukov:2003yp,Davis:2004xb} ($U=-1/X$, $V\propto X$, $F=\rm const.$) and spherically reduced $D$-dimensional Einstein gravity with a Maxwell field \cite{Berger:1972pg, Benguria:1977in, Thomi:1984na, Hajicek:1984mz, Kuchar:1994zk} ($U=-(D-3)/[(D-2)X]$, $V\propto X^{(D-4)/(D-2)}$, $F\propto X$), see table~1 in \cite{Grumiller:2006ja} for a comprehensive list of further models.

Besides intrinsic two-dimensional toy models for quantum gravity, there are numerous applications where a higher-dimensional theory reduces to \eqref{eq:intro0}. In particular, Anti-de~Sitter (AdS) solutions of two-dimensional dilaton gravity are a fairly generic fixed point of extremal black holes through the attractor mechanism \cite{Ferrara:1995ih, Strominger:1996kf, Ferrara:1996dd}, which allows a simple determination of the entropy of such black holes \cite{Sen:2005wa}. AdS$_2$ holography has already a long history, see \cite{Strominger:1998yg, Cadoni:1998sg, Maldacena:1998uz, Spradlin:1999bn, NavarroSalas:1999up, Azeyanagi:2007bj,Chamon:2011xk} for selected earlier references. 

A concrete proposal for AdS$_2$ holography for a specific model \eqref{eq:intro0} was made in \cite{Hartman:2008dq}, who found a non-trivial central charge and $\hat{u}(1)$ level in the anomalous transformation laws for (twisted) stress tensor and current, respectively. Their results were confirmed through a holographic renormalization procedure \cite{Castro:2008ms}. 

In this respect AdS$_2$ holography resembles AdS$_3$ holography, except that there is a single copy of the Virasoro algebra instead of two copies. However, when calculating the canonical boundary charges \'a la Brown and Henneaux \cite{Brown:1986nw} it turns out that they vanish for this particular model in the classical approximation, see for instance appendix A of \cite{Castro:2014ima}, reminiscent of the situation in near horizon extremal Kerr \cite{Amsel:2009ev, Dias:2009ex}. Thus, from an intrinsic two-dimensional perspective there are no physical states, neither in the bulk (like in three dimensions) nor at the boundary (unlike in three dimensions), which is consistent with the analysis of \cite{Balasubramanian:2009bg} and earlier results in \cite{Maldacena:1998uz}. So this part of the story is well-known.

What is not known currently is to what extent these conclusions are specific to the chosen model, the chosen ground state (constant vs.~linear dilaton vacuum), the chosen boundary conditions and/or the classical approximation. Indeed, there exists at least one example of non-trivial AdS$_2$ holography \cite{Grumiller:2013swa}, so clearly there must be some dependence on these choices. Moreover, it was shown in \cite{Vassilevich:2013ai} that any Poisson-sigma model (PSM), which in particular includes dilaton gravity, on a finite cylinder is holographically dual to a noncommutative quantum mechanics. 

The main aim of the present work is to clarify this issue by a comprehensive holographic analysis of all the models described by the bulk action \eqref{eq:intro0} that allow for AdS$_2$ solutions with a constant dilaton field.

The tools used to reach this aim are the gauge theoretic formulation of dilaton gravity \cite{Isler:1989hq, Chamseddine:1989yz, Verlinde:1991rf, Cangemi:1992bj, Ikeda:1993aj, Ikeda:1993fh} as a PSM \cite{Schaller:1994es}, combined with a Brown--Henneaux type of analysis \cite{Brown:1986nw}, and, independently, the Euclidean path integral formulation \cite{Gibbons:1976ue, Gibbons:1994cg}, combined with heat kernel methods \cite{Vassilevich:2003xt}.

Our main conclusion is that AdS$_2$ holography is trivial for constant dilaton boundary conditions for any choices of the functions $U$, $V$  and $F$ in the bulk action \eqref{eq:intro0}. We prove this by determining the canonical charges and, independently, by calculating the quantum gravity partition function. We show the robustness of our conclusion by considering looser boundary conditions, non-linear interactions of the Maxwell field with the dilaton, inclusion of higher spin fields or generic gauge fields.

Therefore, if one would like to study AdS holography in a purely two-dimensional context one has to give up the condition of a constant dilaton. As an example we consider the charged Jackiw--Teitelboim model, whose holographic study was pioneered in \cite{Hartman:2008dq}, and show that it is non-trivial for specific linear dilaton boundary conditions.

The viewpoint taken in this paper is an intrinsic two-dimensional one, without any relations to higher-dimensional theories, since we are interested in genuine AdS$_2$ holography. Naturally, our perspective and scope differ from papers that try to connect with AdS$_3$ holography, see e.g.~\cite{Gupta:2008ki, Castro:2014ima} and Refs.~therein.

This paper is organized as follows.
In section \ref{se:2} we recapitulate basic aspects of two-dimensional dilaton gravity with a Maxwell field and its gauge theoretic formulation, formulate constant dilaton boundary conditions and determine the boundary condition preserving transformations, which lead to a Virasoro algebra with non-vanishing central charge.
In section \ref{se:3} we prove that the canonical charges are trivial and show the robustness of this result by generalizing it in various ways.
In section \ref{se:4} we compute the full quantum gravity partition function and show its triviality.
In section \ref{se:5} we study the charged Jackiw--Teitelboim model with specific linear dilaton boundary conditions, show that it is non-trivial, and determine the entropy macro- and microscopically. 
In section \ref{se:6} we conclude with a discussion.

\section{Two-dimensional dilaton gravity}\label{se:2}

In three dimensions the Chern--Simons formulation \cite{Achucarro:1987vz, Witten:1988hc} of Einstein gravity is extremely useful, particularly for determining the canonical charges \cite{Banados:1998gg}. The analogue in two dimensions is the PSM formulation \cite{Ikeda:1993aj, Ikeda:1993fh, Schaller:1994es} of dilaton gravity, which we review in section \ref{se:2.1}. In section \ref{se:2.2} we formulate constant dilaton boundary conditions and check the transformations that preserve them in section \ref{se:2.3}, where we show the emergence of a Virasoro algebra with non-vanishing central charge.

\subsection{Poisson-sigma model formulation}\label{se:2.1}

Introducing Cartan variables converts the second order action \eqref{eq:intro0} into a first order action that depends on the zweibein $e_a$, the dualized spin-connection\footnote{
We are assuming metric compatibility, $\om^{ab}=-\om^{ba}=\epsilon^{ab}\omega$, but note that all our conclusions generalize to models with non-metricity, since there exists a PSM formulation in that case as well \cite{Adak:2007xc}. 
} $\omega$, the gauge connection $a$, the dilaton $X$, Lagrange multipliers for the torsion constraint $X^a$ and an auxiliary field $f$, which on-shell becomes essentially the electric field $E$. The bulk action is given by
\eq{
I_{\textrm{bulk}} = -\frac{k}{2\pi}\, \int \big( X^a \,(\extd e_a + \epsilon_a{}^b \omega\wedge e_b) + X\extd \omega + f\extd a + \tfrac 12 \epsilon^{ab} e_a \wedge e_b \,{\cal V}(X^c,\,X,\,f) \big)
}{eq:CDV19}
with
\eq{
{\cal V}(X^c,\,X,\,f) = -\tfrac12\,X^a X^b \delta_{ab} \,U(X) - V(X) + f^2/F(X)
}{eq:CDV20}
where the anti-symmetric symbol is defined as $\epsilon^{ab}=\de^a_1 \de^b_0 - \de^a_0 \de^b_1$ and the tangent space metric $\delta_{ab}=\textrm{diag}(1,1)_{ab}$ used to raise and lower Latin indices is chosen with Euclidean signature, in which we shall work from now on. The metric follows in the usual way from the zweibein, $g_{\mu\nu}=e^a_\mu e^b_\nu \delta_{ab}$. The volume form is defined by $\extd^2x\sqrt{g}=\tfrac12\,\epsilon^{ab}e_a\wedge e_b = e_1\wedge e_0 = \ast 1$.

The first order action \eqref{eq:CDV19} can be written as a specific PSM \cite{Schaller:1994es}. The bulk action
\begin{equation}
I_{\textrm{bulk}}=-\frac{k}{2\pi}\, \int \big( X^I\extd A_I + \tfrac 12 P^{IJ}(X^K) A_I \wedge A_J\big)\label{PSM}
\end{equation}
depends on four target space coordinates $X^I=(X,\, X^a,\,f)$, four connection 1-forms $A_X=\om$, $A_a=e_a$, $A_f=a$, and the Poisson tensor 
\begin{equation}
P^{Xb} = X^a\epsilon_{a}^{\ b}\qquad P^{ab} = {\cal V}(X^c,\,X,\,f)\epsilon^{ab}\qquad P^{fX}=P^{fa}=0 \qquad P^{IJ}=-P^{JI}\,.
\label{Ptensor}
\end{equation}
As a consequence of the non-linear Jacobi identities
\eq{
\big(\partial_L P^{IJ}\big)\,P^{LK} + \big(\partial_L P^{JK}\big)\,P^{LI} + \big(\partial_L P^{KI}\big)\,P^{LJ} = 0
}{eq:2d1}
the non-linear gauge transformations
\begin{subequations}
\label{PSMgauge}
\begin{eqnarray}
&&\delta_\lambda X^I=P^{IJ}\lambda_J\\
&&\delta_\lambda A_I=-\extd\lambda_I -\partial_I P^{JK} \lambda_K A_J
\end{eqnarray}
\end{subequations}
leave the PSM action (\ref{PSM}) invariant up to a total derivative,
\begin{equation}
\delta_\lambda S=\frac{k}{2\pi}\, \int\extd \big[A_I\lambda_J \big(X^K \partial_K P^{IJ} - P^{IJ} \big) \big]\,.\label{totder}
\end{equation}

The equations of motion (EOM)
\begin{subequations}
 \label{eq:CDV7}
\begin{align}
 \extd X^I + P^{IJ}A_J &= 0 \label{eq:EOM1}\\
 \extd A_I + \tfrac12\,\partial_I P^{JK} A_J\wedge A_K &= 0 \label{eq:EOM2}
\end{align}
\end{subequations}
have two classes of solutions: linear and constant dilaton vacua (CDVs). We are mostly interested in the latter.

In PSM language linear dilaton vacua have two Casimir functions, the mass and charge, which parametrize the space of solutions \cite{Grosse:1992vc, Mann:1992yv}. The Poisson tensor then has rank 2 on-shell  \cite{Schaller:1994es}. By contrast, CDVs have four Casimir functions, and the Poisson tensor has rank 0 on-shell. In this case the Casimir functions are simply the target space coordinates,
\eq{
X^I = \bar X^I = \textrm{const.}\qquad P^{IJ}(\bar X^k) = 0\,.
}{eq:CDV8}
For the Poisson tensor \eqref{Ptensor} we get
\eq{
X^c = 0 \qquad X = \bar X\qquad  f = \bar f
}{eq:CDV6}
where the constants  $\bar X$ and $\bar f$ are related by the condition
\eq{
{\cal V}(X^c=0,\,X=\bar X,\,f=\bar f) = 0\,.
}{eq:CDV4}
The EOM \eqref{eq:EOM1} then hold trivially.
The EOM \eqref{eq:EOM2} for the connection 1-forms $A_I$ imply vanishing torsion $T_a$, constant curvature $R$ and constant electric field $E$.
\begin{subequations}
 \label{eq:CDV21}
\begin{align}
 \extd e_a + \epsilon_a{}^b\,\om\wedge e_b = 0 &= T_a \\
 \ast\extd \om = -\partial_X{\cal V} &= \tfrac12\,R \label{eq:R} \\
 \ast\extd a = -\partial_f{\cal V} &= -E \label{eq:E}
\end{align}
\end{subequations}
To get an AdS$_2$ solution with unit AdS radius, $R=-2$, we additionally demand 
\eq{
\partial_X {\cal V}(X^c,\,X,\,f)\big|_{X^c=0,\,X=\bar X,\,f=\bar f} = 1\,.
}{eq:CDV5}
Since ${\cal V}$ has a dimension of inverse length squared, the condition \eqref{eq:CDV5} always can be achieved by a rescaling of units, provided the quantity $\partial_X {\cal V}$ is positive on a given CDV, which is necessarily the case for AdS vacua.

\subsection{Constant dilaton boundary conditions}\label{se:2.2}

The line-element for AdS$_2$ CDVs with unit AdS radius in Fefferman--Graham gauge reads 
\eq{
\extd s^2 = \extd\rho^2 + \tfrac14\,\big(e^{2\rho} - 2M(\varphi) + M^2(\vp)e^{-2\rho} \big)\,\extd\varphi^2
}{eq:CDV22} 
where we assume periodicity in Euclidean time, $\varphi\to\varphi+\beta$. In Schwarzschild gauge the line-element is given by
\eq{
\extd s^2 = \big(r^2-M(\vp)\big)\,\extd\vp^2 + \frac{\extd r^2}{r^2-M(\vp)}\,.
}{eq:CDV23}
For zero mode solutions, $M=M_0 > 0$, there is a Killing horizon at $r=\sqrt{M_0}$ with Hawking temperature $T=\sqrt{M_0}/(2\pi)$. Given some periodicity $\beta=2\pi/\sqrt{M_0}$, there are exactly two smooth solutions. 
\begin{align}
 \textrm{Global\;}\mathbb{H}^2: & \extd s^2 = \extd\rho^2 + \cosh^2\!\rho\,\extd\varphi^2 \label{eq:global}\\
 \textrm{Poincar\'e\;}\mathbb{H}^2: & \extd s^2 = \extd\rho^2 + \frac14\, \Big(e^\rho - \frac{4\pi^2}{\beta^2}\,e^{-\rho}\Big)^2\,\extd\varphi^2 \label{eq:poincare}
\end{align}
For $\beta=2\pi$ the second solution simplifies to $\extd s^2 = \extd\rho^2 + \sinh^2\!\rho\,\extd\varphi^2$, which is the metric of the Lobachevsky plane $\mathbb{H}^2$.

We define CDV boundary conditions in the PSM formulation that lead asymptotically to the above solutions for the line-element and constant dilaton and electric field.
\begin{subequations}
 \label{eq:CDV1}
\begin{align}
 X^0 &= 0 & e_{\varphi 0} &= \tfrac12\,e^\rho - \tfrac12\,e^{-\rho} M(\varphi) + {\cal O}(e^{-3\rho}) & e_{\rho 0} &= 0 \\
 X^1 &= 0 & e_{\varphi 1} &= 0 & e_{\rho 1} &= 1 \\
 X &= \bar X & \omega_\varphi &= -\tfrac12\,e^\rho - \tfrac12\,e^{-\rho} M(\varphi) + {\cal O}(e^{-3\rho}) & \omega_\rho &= 0 \\
 f &= \bar f & a_\varphi &= E\,\om_\vp + j(\vp) + {\cal O}(e^{-2\rho}) & a_\rho &= 0
\end{align}
\end{subequations}
The constants $\bar X$ and $\bar f$ are related through the conditions \eqref{eq:CDV4} and the electric field $E$ is given by \eqref{eq:E}.
We have gauge-fixed as much as possible, using Fefferman--Graham gauge for the zweibein and axial gauge for spin- and gauge-connections. These are essentially the same boundary conditions as used in \cite{Castro:2008ms}, reformulated in PSM language and generalized to arbitrary dilaton gravity models \eqref{eq:intro0}. There are two free functions of the Euclidean time $\vp$ appearing in our boundary conditions, $M(\vp)$ and $j(\vp)$. They are candidates for canonical boundary charges.

\subsection{Boundary condition preserving transformations}\label{se:2.3}

We consider now all transformations \eqref{PSMgauge} that preserve the gauge and boundary conditions \eqref{eq:CDV1} and find 
\begin{subequations}
 \label{eq:CDV11}
\begin{align}
 \la_0 &= \tfrac12\,\la(\vp) e^{\rho} - \big(\tfrac12\,\la(\vp)M(\vp)+\la''(\vp)\big)\, e^{-\rho} \\
 \la_1 &= -\la^\prime(\vp) \\
 \la_X &= -\tfrac12\,\la(\vp) e^{\rho} - \big(\tfrac12\,\la(\vp)M(\vp)+\la''(\vp)\big)\, e^{-\rho} \\
 \la_f &= E\,\la_X + \mu(\vp) \,.
\end{align}
\end{subequations}
Thus, we have two free functions, $\la(\vp)$ and $\mu(\vp)$, parametrizing all allowed boundary condition preserving transformations. 

Gauge transformations \eqref{PSMgauge} with gauge parameters \eqref{eq:CDV11} yield the transformations laws for the free functions $M$ and $j$:
\begin{align} 
 \de M &= - M^\prime\la - 2M \la^\prime - 2 \la''' \label{eq:CDV24}\\
 \de j &= -\mu' \label{eq:CDV25} 
\end{align}
These results are compatible with the ones in \cite{Hartman:2008dq, Castro:2008ms, Castro:2014ima}, but now are valid for arbitrary dilaton gravity models with an AdS CDV. In particular, the presence of a Maxwell field is in no way essential for the appearance of the infinitesimal Schwarzian derivative in \eqref{eq:CDV24}.

Introducing the normalization factor $\alpha$ for the Virasoro zero mode, $L_0 = \alpha M_0$, we find that our result \eqref{eq:CDV24} is compatible with the assumption that the asymptotic symmetry algebra contains a Virasoro algebra $[L_n,\,L_m]=(n-m)\,L_{n+m}+c/12\,(n^3-n)\,\de_{n+m,\,0}$ with central charge
\eq{
c = 24\alpha\,.
}{eq:CDV26} 
The chiral Cardy formula would then yield an entropy
\eq{
S_{\textrm{\tiny Cardy}} = \frac{\pi^2 c\, T}{3} = 2\pi \sqrt{\frac{c L_0}{6}} = 4\pi\alpha\sqrt{M_0}\,.
}{eq:CDV27}
However, we have not checked yet whether there is a non-trivial asymptotic symmetry algebra in the first place; it could be that the transformations \eqref{eq:CDV24}, \eqref{eq:CDV25} are pure gauge, in which case the theory would contain no physical states besides the vacuum. In order to decide this important issue we construct the canonical charges in the next section.

\section{Canonical charges}\label{se:3}

In this section we determine the asymptotic symmetry algebra, which consists of all boundary condition preserving transformations \eqref{eq:CDV11}, modulo trivial gauge transformations. The canonical charges allow to decide which transformations are non-trivial and which are pure gauge. 

In section \ref{se:3.1} we construct the canonical charges and prove that they are trivial, so that trivial gauge transformations exhaust all boundary condition preserving transformations \eqref{eq:CDV11}, which implies that the asymptotic symmetry algebra is empty. 
In section \ref{se:3.2} we show that making the boundary conditions looser does not change these conclusions. 
In section \ref{se:3.3} we generalize the triviality of AdS$_2$ holography for CDVs to theories that contain more complicated interactions with the Maxwell-field and/or higher spin fields and/or non-abelian gauge fields.

\subsection{Triviality of canonical charges}\label{se:3.1}

A straightforward canonical analysis \cite{Kummer:1996hy} together with the Castellani algorithm \cite{Castellani:1981us} yields the canonical boundary currents \cite{Grumiller:2013swa}
\eq{
\delta Q[\la_I] = \frac{k}{2\pi}\,\delta X^I\,\la_I\big|_{\rho\to\infty}\,.
}{eq:CDV28}
The key observation in deriving the result \eqref{eq:CDV28} is that the secondary first class constraints contain spatial derivatives of the target space coordinates \cite{Kummer:1996hy}. There is actually a shortcut to see the result \eqref{eq:CDV28} directly from the action, which we now make explicit.

The full action (corresponding to ``simplest boundary conditions'' in the discussion of appendix B.1 in \cite{Bergamin:2004us}) consists of the bulk action \eqref{PSM} and a boundary action that effectively partially integrates the first term:
\eq{
\Gamma = I_{\textrm{\tiny bulk}} + \frac{k}{2\pi}\,\int_{\partial{\cal M}} X^IA_I = - \frac{k}{2\pi}\, \int \big(A_I\wedge\extd X^I + \tfrac 12 P^{IJ}(X^K) A_I \wedge A_J\big)\,.
}{eq:CDV29}
The action \eqref{eq:CDV29} vanishes if evaluated on CDVs
\eq{
\Gamma\big|_{\textrm{\tiny CDV}} = 0
}{eq:CDV30}
and has a well-defined variational principle for our boundary conditions \eqref{eq:CDV1}
\eq{
\delta\Gamma\big|_{\textrm{\tiny CDV}} = \frac{k}{2\pi}\,\int_{\partial \cal M} A_I \, \delta X^I = 0
}{eq:CDV31}
since $\de X^I=0$. Off-shell, the gauge variation of the action \eqref{eq:CDV29} yields a boundary term
\eq{
\delta_\la \Gamma = \frac{k}{2\pi}\,\int_{\partial\cal M} X^I\extd\lambda_I
}{eq:CDV32}
which vanishes on CDVs upon partial integration. Partial integration introduces a boundary term $B$ at the intersection $\partial{\cal M} \cap \Sigma$ where $\Sigma$ is a constant time hypersurface,
\eq{
B[\lambda_I] = \frac{k}{2\pi}\,X^I\lambda_I\big|_{\rho\to\infty}
}{eq:CDV33}
whose variation in field space coincides precisely with the canonical currents \eqref{eq:CDV28}. 

The canonical currents \eqref{eq:CDV28} vanish identically for our boundary conditions \eqref{eq:CDV1}, and therefore the canonical charges are state-independent and hence trivial. This means that the asymptotic symmetry algebra is empty, and the boundary condition preserving transformations \eqref{eq:CDV24}, \eqref{eq:CDV25} are pure gauge. 

Another interesting property of the canonical currents \eqref{eq:CDV28} is their gauge invariance
\eq{
\de_{\la^2_J} Q[\la^1_I] = \frac{k}{2\pi}\,P^{IJ}\la_J^2\,\la^1_I\big|_{\rho\to\infty} = 0
}{eq:CDV34}
due to the CDV conditions \eqref{eq:CDV8}. This shows that even non-infinitesimal transformations (connected with the identity) cannot make the canonical currents non-trivial. This is different from the situation encountered for linear dilaton solutions \cite{Gegenberg:1997de}, where boundary states were found in this way.

In the next subsections we try --- and fail --- to circumvent these conclusions by considering looser boundary conditions and more complicated interactions, which shows the robustness of our conclusion that AdS$_2$ holography is trivial for CDVs. We stress again that this result is independent of the presence or absence of a Maxwell field, as long as an AdS CDV exists.

\subsection{Looser boundary conditions}\label{se:3.2}

Perhaps the boundary conditions \eqref{eq:CDV1} are simply too strict. Indeed, we have switched off all fluctuations of the target space coordinates, but we could have allowed instead some asymptotic fall-off. From the canonical currents \eqref{eq:CDV28} we see that a fall-off behavior of the dilaton field of the form $\delta X = {\cal O}(e^{-\rho})$ could produce finite canonical charges, since the gauge parameter $\la_X$  in \eqref{eq:CDV11} diverges like $e^\rho$. Motivated by this observation we consider now looser boundary conditions that allow for such terms.
\begin{subequations}
 \label{eq:CDV14}
\begin{align}
 X^0 &= X^0_{(1)}(\vp) e^{-\rho} + {\cal O}(e^{-2\rho}) \\
 X^1 &= X^1_{(1)}(\vp) e^{-\rho} + {\cal O}(e^{-2\rho}) \\
 X &= \bar X + X_{(1)}(\vp) e^{-\rho} + {\cal O}(e^{-2\rho}) \\
 f &= \bar f + f_{(1)}(\vp) e^{-\rho} + {\cal O}(e^{-2\rho})  \\
 e_{\varphi 0} &= \tfrac12\,e^\rho + e_{\varphi 0}^{(0)}(\vp) + e_{\varphi 0}^{(1)}(\vp) e^{-\rho} + {\cal O}(e^{-2\rho}) \\
 e_{\rho 0} &= e_{\rho 0}^{(1)}(\vp) e^{-\rho} + {\cal O}(e^{-2\rho}) \\
 e_{\varphi 1} &= e_{\vp 1}^{(0)}(\vp) + e_{\vp 1}^{(1)}(\vp) e^{-\rho} + {\cal O}(e^{-2\rho}) \\
 e_{\rho 1} &= 1 +  e_{\rho 0}^{(1)}(\vp) e^{-\rho} + {\cal O}(e^{-2\rho}) \\
 \omega_\varphi &= -\tfrac12\,e^\rho + \om_\vp^{(0)}(\vp) + \om_\vp^{(1)}(\vp) e^{-\rho} + {\cal O}(e^{-2\rho}) \\
 \omega_\rho &= \om_\rho^{(1)}(\vp) e^{-\rho} + {\cal O}(e^{-2\rho}) \\
 a_\varphi &= E\,\om_\vp + a_\vp^{(0)}(\vp) + a_\vp^{(1)}(\vp) e^{-\rho} + {\cal O}(e^{-2\rho})\\
 a_\rho &= a_\rho^{(1)}(\vp) e^{-\rho} + {\cal O}(e^{-2\rho})
\end{align}
\end{subequations}
Again, the constants $\bar X$ and $\bar f$ are related through the conditions \eqref{eq:CDV4} and the electric field $E$ is given by \eqref{eq:E}. Note that we particularly allow for fluctuations
\eq{
\de X^I = {\cal O}(e^{-\rho})\,.
}{eq:CDV63}

Since we are mostly interested in the evaluation of the canonical currents \eqref{eq:CDV28}, we impose on-shell conditions on all the fluctuation terms that we have written explicitly. The EOM impose the conditions
\begin{align}
X_{(1)} &= X_{(1)}^0 \\
f_{(1)} &= 0  %f_{(2)} = 0 %\\
%e_{\vp 0}^{(0)} &= -\om_\vp^{(0)} = \tfrac12\, e_{\rho 1}^{(1)} \\
%e_{\rho 0}^{(1)} &= -\om_\rho^{(1)} \\
%X_{(1)}\partial_X^2 {\cal V} &= 0\,.
\end{align}
on the subleading components of the target space coordinates. There are further restrictions on the functions appearing in the loose boundary conditions \eqref{eq:CDV14}, but we do not need them for our conclusions. Note that the conditions above imply
\eq{
\delta X^0 = \delta X + {\cal O}(e^{-2\rho})\qquad \delta f = {\cal O}(e^{-2\rho}) \,.
}{eq:CDV36}

The gauge parameters that preserve the boundary conditions \eqref{eq:CDV14} can be similarly expanded
\begin{subequations}
 \label{eq:CDV15}
\begin{align}
 \la_0 &= \tfrac12\,\la(\vp) e^{\rho} + \la_0^{(0)} + \la_0^{(1)} e^{-\rho} + {\cal O}(e^{-2\rho}) \\
 \la_1 &= \la_1^{(0)} + \la_1^{(1)} e^{-\rho} + {\cal O}(e^{-2\rho}) \\
 \la_X &= -\tfrac12\,\la(\vp) e^{\rho} + \la_X^{(0)} + \la_X^{(1)} e^{-\rho} + {\cal O}(e^{-2\rho}) \\
 \la_f &= E\,\la_X + \mu(\vp) + \la_f^{(1)} e^{-\rho} + {\cal O}(e^{-2\rho}) \,.
\end{align}
\end{subequations}
Again, there will be restrictions on the functions appearing in the gauge parameters \eqref{eq:CDV15}, and again we do not need them for our conclusions.

With the boundary conditions and gauge parameters above the canonical currents \eqref{eq:CDV28} expand to a sum of order unity terms [due to \eqref{eq:CDV63}] and subleading terms
\eq{
\de Q[\la_I] = \frac{k\la(\vp)}{4\pi}\,e^{\rho_c} \big(\de X^0 - \de X - E\,\de f\big) + {\cal O}(e^{-\rho_c})\,.
}{eq:CDV35}
Here $\rho_c\gg 1$ is the cut-off surface where the charges are evaluated. Taking the cut-off to infinity, $\rho_c\to\infty$, removes the subleading terms ${\cal O}(e^{-\rho_c})$.
However, due to the relations \eqref{eq:CDV36} the order unity terms cancel precisely and the canonical currents vanish.

Therefore, even for the looser set of boundary conditions \eqref{eq:CDV14} the canonical charges are trivial.
A further generalization is achieved by introducing a chemical potential, which effectively amounts to replacing the factor $1/2$ in the leading terms in $e_{\vp 0}$, $\om_\vp$ and $a_\vp$ by some function of $\vp$. We have checked that the conclusions remain unchanged: the canonical charges are trivial. The theory has only one physical state, the vacuum.

Let us discuss one final generalization of boundary conditions. We can allow $X$ and $f$ to fluctuate to ${\cal O}(1)$, as long as the condition \eqref{eq:CDV4} remains intact. This modifies the previous boundary conditions by making $\bar X$ and $\bar f$ state dependent, so that the following fluctuations are allowed additionally
\eq{
\de X = -E\,\de f \qquad \de f = {\cal O}(1)\,.
}{eq:CDV64}
In this case there is a non-trivial, integrable and finite $U(1)$ charge.
\eq{
Q[\mu] = \frac{k}{2\pi}\, f \mu(\vp)
}{eq:CDV65}
However, there are still no diffeomorphism charges, and the asymptotic symmetry algebra is trivial, since any gauge variation of the charge \eqref{eq:CDV65} vanishes due to $\de_{\la_I} f = P^{fI}\la_I = 0$. We consider this case as somewhat artificial, as the boundary electric field \eqref{eq:E} is allowed to vary.

In what follows we generalize our result of triviality of the canonical charges to theories with additional interactions.

\subsection{More general actions}\label{se:3.3}

We consider now more general actions that lead to the same conclusions. The simplest generalization is to replace the function ${\cal V}$ defined in \eqref{eq:CDV20} by
\eq{
{\cal V}(X^c,\,X,\,f) = -\sum_{n=1}^\infty (\tfrac12\,X^a X^b\delta_{ab})^n U_n(X) - V(X,\,f)
}{eq:CDV37}
which includes now also models like dimensionally reduced conformal gravity \cite{Guralnik:2003we, Grumiller:2003ad} and models where the Maxwell field interacts arbitrarily non-linearly with the dilaton.
None of our conclusions is changed by this generalization, as long as solutions of \eqref{eq:CDV4} exist [with the normalization condition \eqref{eq:CDV5} that sets the AdS radius to unity].

Higher spin theories in two dimensions can also be formulated as a PSM \cite{Alkalaev:2013fsa, Grumiller:2013swa}. The Poisson tensor takes the form $P^{IJ}=f^{IJ}{}_K X^K$, where $f^{IJ}{}_K$ are structure constants of some Lie algebra that contains $sl(2)$ \cite{Grumiller:2013swa}. For CDVs the Poisson tensor vanishes on-shell, $P^{IJ}=0$, and we get again as many Casimir functions as there are target space coordinates. Then the same features as in the spin-2 case are encountered. This is seen most easily in the analogue of the stricter boundary conditions \eqref{eq:CDV1}. As long as all fluctuations $\delta X^I$ are switched off the canonical charges \eqref{eq:CDV28} must vanish identically. For looser boundary conditions analogue to \eqref{eq:CDV14} the EOM \eqref{eq:EOM1} constrain again the relevant fluctuations similar to \eqref{eq:CDV36} so that the canonical boundary charges are again trivial. 

The generalization to Yang--Mills theories follows a similar logic. Yang--Mills theories interacting with dilaton gravity can again be formulated as a PSM, see for instance \cite{Strobl:1999wv} and references therein. The difference to higher spin theories is that the Poisson tensor now does not have to vanish identically for a CDV. Instead, it is sufficient if the gravity part of the Poisson tensor behaves as described in sections \ref{se:2} and \ref{se:3}, while the Yang--Mills part takes the form $P^{ij}=f^{ij}{}_k X^k \neq 0$. This is the main difference to the higher spin case. So now we have a number of Casimirs that depends on the Yang--Mills gauge algebra, plus the three `gravitational Casimirs', $X$ and $X^a$. The main point here is that there is essentially no mixing between the gravity and the Yang--Mills part. The only appearance of Yang--Mills fields in the gravity part is through the Yang--Mills Casimirs, which enter the generalization of the potential $\cal V$ defined in \eqref{Ptensor} for the $u(1)$ case; the dependence on $f$ is simply replaced by a dependence on all possible Yang--Mills Casimirs. The discussion of boundary conditions and canonical charges is then completely analogous to the $u(1)$ case. In particular we recover again the result that the diffeomorphism charges are trivial.

In all the examples so far we have seen that the canonical diffeomorphism charges are trivial classically. In the next section we check indirectly if this statement also holds at the quantum level by calculating the full quantum gravity partition function.

\section{Quantum gravity partition function}\label{se:4}

The canonical analysis of the previous section is classical. It is conceivable that switching on quantum effects makes the theory non-trivial. After all, the asymptotic symmetry algebra and the canonical charges could receive quantum corrections, so even if the classical results show triviality the quantum mechanical results might be non-trivial. In this section we rule out this possibility by considering the full quantum gravity partition function and showing that it is unity.

In section \ref{se:4.1} we determine the classical contribution.
In section \ref{se:4.2} we calculate the one-loop partition function and argue that the theory should be one-loop exact.
In section \ref{se:4.3} we determine all instanton corrections and collect the results.

\subsection{Classical partition function}\label{se:4.1}

We use the Euclidean path integral formulation \cite{Gibbons:1976ue, Gibbons:1994cg}. Our aim is to determine the full quantum gravity partition function
\eq{
Z = \int_{\textrm{\tiny bc}}({\cal D}X^I)({\cal D}A_I)({\textrm{measure}})\,\exp{\big(-\Gamma[X^I,\,A_I]\big)}
}{eq:CDV56}
where `bc' denotes that we evaluate the path integral for certain boundary and smoothness conditions, `measure' refers to the ghost- and gauge-fixing part, and $\Gamma$ is the full action \eqref{eq:CDV29}. Results for the exact path integral have shown quantum triviality, i.e., the quantum partition function equals the classical one \cite{Kummer:1996hy}. However, the previous calculations did not take into account asymptotic boundary conditions, nor possible global effects, nor instanton contributions. This is why we re-evaluate the path integral. As we shall see, the local results of \cite{Kummer:1996hy} are not modified globally for CDVs.

We make now an expansion of the path-integral into classical contribution (c), perturbative corrections (p) and non-perturbative corrections (n).
\eq{
Z = Z_{\rm c} \times Z_{\rm p} \times Z_{\rm n}
}{eq:CDV57}
We start with the classical piece.
\eq{
Z_{\rm c} = \exp{\big(-\Gamma|_{\textrm{\tiny CDV}}\big)}
}{eq:CDV58}
With the result \eqref{eq:CDV30} we then obtain
\eq{
Z_{\rm c} = 1\,.
}{eq:CDV59}
Thus, the classical partition function is trivial, which concurs of course with the conclusions of section \ref{se:3} that the canonical charges are trivial.

\subsection{Perturbative corrections}\label{se:4.2}

Let us consider now the perturbative corrections $Z_{\rm p}$ to the classical partition function \eqref{eq:CDV59}. Given that our theory is a topological field theory of Schwarz type \cite{Birmingham:1991ty}, namely a PSM \cite{Schaller:1994es}, one can argue that the theory should be one-loop exact, either along similar lines as \cite{Maloney:2007ud} (who applied this to 3-dimensional gravity) or using one-loop exactness of anomalies, see e.g.~\cite{Bertlmann:1996xk}, and arguing that the perturbative corrections should be determined uniquely by the anomalies. We assume that there is no relevant subtlety with these arguments, so that the one-loop partition function captures the full information about all perturbative corrections. We shall briefly come back to this issue below while performing the one-loop calculation.

In the one-loop calculation we use bars to denote classical values, while un-barred quantities will be quantum fluctuations. The action quadratic in quantum fluctuations reads
\begin{equation}
S_2= -\frac k{2\pi} \int\extd^2x \left[ \tilde\epsilon^{\mu\nu} X^I (\partial_\mu A_{\nu I} +\Omega_{\mu I}{}^JA_{\nu J} )+
(\bar e) \frac 12\frac {\partial^2 \mathcal{V}(\bar X^K)}{\partial X^I\partial^K} X^IX^K \right] \label{S2}
\end{equation}
where $(\bar e)=\tfrac 12 \tilde \epsilon^{\mu\nu}\epsilon^{ab}\bar e_{\mu a}\bar e_{\nu b}$ and
\begin{eqnarray}
&& \Omega_{\mu a}{}^b=\bar \omega_\mu \epsilon_a^{\ b}\\
&& \Omega_{\mu a}{}^X=-\epsilon_{a}^{\ b} \bar e_{\mu b}\qquad 
\Omega_{\mu X}{}^a=-\epsilon^{ab} \bar e_{\mu b}\label{OmuXa}\\
&& \Omega_{\mu f}{}^a= -\bar E \epsilon^{ab} \bar e_{\mu b}\,.
\end{eqnarray} 
All other components of the connection $\Omega$ vanish. To derive (\ref{S2}) we used that the classical fields satisfy (\ref{eq:CDV6})-(\ref{eq:CDV5}) for an AdS$_2$ CDV with unit AdS radius.

The same connection appears in the linearized gauge transformations
\begin{equation}
\delta_\lambda X^J=0\qquad \delta_\lambda A_{\mu I}=-\nabla_\mu \lambda_I=
-\partial_\mu \lambda_I -\Omega_{\mu I}{}^J\lambda_J \,.\label{gautr}
\end{equation}
The invariance of the quadratic action (\ref{S2}) under gauge transformations (\ref{gautr}) implies that the connection $\Omega$ is flat.
\begin{equation}
[\nabla_\mu,\nabla_\nu]=\partial_\mu \Omega_\nu - \partial_\nu \Omega_\nu + [\Omega_\mu,\Omega_\nu ]=0
\label{flat}
\end{equation}
The flatness of the connection can also be verified by direct calculation.

Let us expand the fluctuations $A_{\mu I}$ into a sum of gauge ($\la$) and transverse ($\chi$) parts.
\begin{equation}
A_{\mu I}=-\nabla_\mu \lambda_I +\varepsilon_\mu^{\ \nu}\nabla_\nu^\dag \chi_I +A_{\mu I}^{(\rm h)}\label{expan}
\end{equation}
Here $\varepsilon_{\mu}^{\ \nu}$ is the Levi-Civit\'a tensor. $A_{\mu I}^{(\rm h)}$ correspond to square integrable (twisted) harmonic
one-forms, that are both longitudinal and transverse and are given by gradients (with $\nabla_\mu$) of non-normalizable zero
modes of the scalar operator $\nabla_\mu^\dag \nabla_\mu$.  As argued in \cite{Larsen:2014bqa}, the harmonic one-forms correspond
to boundary modes of the theory. Interestingly, these modes do not contribute to the quadratic action (\ref{S2}). This facts hints
to the holographic triviality of CDVs. This is in contrast to the model considered in \cite{Bertin:2012qw} where non-integrable 
scalar modes on the hyperbolic plane generate physical boundary states.

The presence of infinitely many harmonic one-forms complicates computations of the partition function on $\mathbb{H}^2$. To avoid this difficulty we use the method of continuation of the partition function from the two-sphere $S^2$ \cite{Larsen:2014bqa,Gupta:2013sva}. 
The unit $S^2$ is a CDV corresponding to the zeros of $\mathcal{V}(X,X^a,f)$ where $\partial_X\mathcal{V}=-1$ instead of $+1$ in Eq.\ 
(\ref{eq:CDV5}). Nonvanishing components of the zweibein and spin-connection read: $\bar e_{\rho 1}=1$, $\bar e_{\varphi 0}=\sin (\rho)$,
$\bar \omega_\varphi = -\cos (\rho)$. The only modification of the connection $\Omega_\mu$ is the sign flip of $\Omega_{\mu X}{}^a$, that becomes
\begin{equation}
\Omega_{\mu X}{}^a=-\epsilon^{ab} \bar e_{\mu b}
\end{equation}
on $S^2$.

We define the path integral measure ${\mathcal{D}}A_{\mu J}$ by the identity
\begin{equation}
\int \mathcal{D}A_{\mu J}\, e^{-\langle A,A \rangle }=1 \,.\label{pint}
\end{equation}
The path integral measure is, therefore, defined by the inner product $\langle \ ,\ \rangle$. We take an
ultralocal product
\begin{equation}
\langle A,A' \rangle = \int\extd^2x (\bar e) \delta^{IJ} \bar g^{\mu\nu} A_{\mu I}A'_{\nu J} 
\label{prod}
\end{equation}

The connection $\Omega$ is not hermitian with respect to the inner product (\ref{prod}). This implies in
particular that $\nabla^\dag \ne -\nabla$. However, it can be transformed to a hermitian one,
\begin{equation}
\nabla_\mu = \Phi^{-1} \hat \nabla_\mu \Phi\qquad \hat\nabla = - \hat\nabla^\dag
\end{equation}
with the field 
\begin{equation}
\Phi={\rm Id}+\phi\qquad \phi_f{}^X=-\bar E\label{Phi}
\end{equation}
where Id is the identity. Other components of $\phi$ vanish so that the inverse of $\Phi$ is given by $\Phi^{-1}={\rm Id}-\phi$. Furthermore, the hermitian derivative $\hat\nabla$ can be decomposed as
\begin{equation}
\hat\nabla_\mu =\Psi^{-1} \partial_\mu\, \Psi\,.  \label{l1l2}
\end{equation}
The gauge transformation $\Psi$ is unitary with respect to the inner product
defined above. It reads
$\Psi=\exp (\varphi l_2)\, \exp(\rho l_1)$
where the only non-zero matrix elements of $l_1$ and $l_2$ are $(l_1)_0{}^X=-(l_1)_X{}^0=(l_2)_1{}^0=-(l_2)_0{}^1=1$. 

There are no twisted harmonic one-forms on $S^2$,
\begin{equation}
A_{\mu I}^{({\rm h})}=0\,.
\end{equation}
We postpone the demonstration of this fact until after Eq.\ (\ref{Trt}).
 Due to the flatness of the connection (\ref{flat}) this decomposition is orthogonal with respect to the inner product (\ref{prod}). 

The change of variables $A_{\mu J}\to \lambda_J,\ \chi_J$ induces a Jacobian factor, 
$\mathcal{D}A_{\mu J}=\mathcal{J}\, \mathcal{D} \lambda_J\mathcal{D} \chi_J$,
which can be easily found by substituting the decomposition (\ref{expan}) in the definition of the measure (\ref{pint}) and performing Gaussian integrals over $\lambda$ and $\chi$. This yields the Jacobian
\begin{equation}
\mathcal{J}=\det (\nabla_\mu^\dag \nabla^\mu)^{\frac 12} \cdot \det (\nabla_\mu \nabla^{\mu\dag)})^{\frac 12}\label{J}\,.
\end{equation}

The one-loop partition function then decomposes into path integrals over $X$, $\la$ and $\chi$.
\begin{align}
Z&=\int \mathcal{D}X\, \mathcal{D} A\, \exp(-S_2) \nonumber\\
&=\int \mathcal{D}X\, \mathcal{D} \lambda \, \mathcal{D} \chi \, \mathcal{J}
\exp\left[ \frac k{2\pi} \int d^2x (\bar e) \left( -X^I \nabla^\mu \nabla_\mu^\dag \chi_I
+\frac 12\frac {\partial^2 \mathcal{V}(\bar X^K)}{\partial X^I\partial X^K} X^IX^K \right) \right] 
\end{align}
The integration over $\lambda$ is performed trivially, yielding an infinite volume of the gauge group, which we discard.
The integration over $X^I$ and $\chi_I$ gives
\begin{equation}
Z=\mathcal{J} \cdot \det( \nabla^\mu \nabla_\mu^\dag)^{-1}=
\frac {\det (\nabla_\mu^\dag \nabla^\mu)^{\frac 12}}{
\det (\nabla_\mu \nabla^{\mu\dag})^{\frac 12}} \,.\label{Z}
\end{equation}
Interestingly, the terms in $S_2$ that are quadratic in fluctuations of the target space coordinates $X^I$ have no influence on the partition function. This means that our results are universal for AdS$_2$ CDVs, regardless of the specific properties of the potentials in the action. Obviously, for $\bar E=0$ we have $\nabla=-\nabla^\dag$, and the partition function is trivial, $Z=1$. This means we have proven that the one-loop partition function is trivial if the electric field vanishes. 

The only reason why the partition function (\ref{Z}) has a chance to be non-trivial is that the transformation $\Phi$ is not unitary, like in the case of conformal and chiral transformations. The structure of variation of $Z$ with respect to $\Phi$ reminds very much of more conventional anomalies. E.g., this variation is given by the heat kernel coefficients, as we shall see below. Basing on this analogy with anomalies, we conjecture that the partition function has no higher loop corrections, see the paragraph at the beginning of section \ref{se:4.2}.

To evaluate $Z$ we shall use the methods developed earlier in Refs.\ \cite{Vassilevich:2000kt,Gilkey:2002qd}. It is convenient to make a polar decomposition of the matrix $\Phi$ as $\Phi=HU$, with a hermitian matrix $H$ and a unitary matrix $U$.
\begin{equation}
H=\left( \begin{array}{cccc} 1 & 0 & 0 & 0 \\
0 & 1 & 0 & 0 \\
0 & 0 &\cos (\alpha) & -\sin (\alpha) \\
0 & 0 &-\sin (\alpha) & \frac 2{\cos(\alpha)} -\cos(\alpha) \\
\end{array}\right) \qquad
U=\left( \begin{array}{cccc} 1 & 0 & 0 & 0 \\
0 & 1 & 0 & 0 \\
0 & 0 & \cos(\alpha) & \sin(\alpha) \\
0 & 0 & -\sin(\alpha) & \cos(\alpha)
\end{array}\right)
\end{equation}
Here $\alpha={\rm arctan}\, (\bar E/2)$. The partition function \eqref{Z} then only depends on the hermitian matrix $H$.
\begin{equation}
Z=
\frac {\det (-H \hat \nabla_\mu H^{-2} \hat \nabla^\mu H)^{\frac 12}}{
\det (-H^{-1} \hat \nabla_\mu H^2 \hat\nabla^{\mu}H^{-1})^{\frac 12}} \label{Z2}
\end{equation}
Let $\delta_\alpha H$ be the variation of $H$ with respect to $\alpha$. Then
\begin{equation}
\delta_\alpha H^2 = H \mat H \qquad \delta_\alpha H^{-2} =-H^{-1}\mat H^{-1} \label{varH2}
\end{equation}
where
\begin{equation}
\mat = H^{-1}\cdot \delta_\alpha H + \delta_\alpha H \cdot H^{-1}\,.\label{beta}
\end{equation}
It is easy to check that the matrix $\mat$ is traceless.
\begin{equation}
{\rm tr}\, \mat=0\label{trb}
\end{equation}

Consider the space $\mathbb{R}^4 \otimes \Lambda {S}^2$ of differential forms that also carry a target space index $I$. Let us take the usual exterior derivative $\del$ and coderivative $\updelta$ on $ \Lambda {S}^2$ and twist them.
\begin{equation}
\del_H=H^{-1}\Psi^{-1} \del \Psi H\qquad \updelta_H = H \Psi^{-1} \updelta \Psi H^{-1} \label{twist}
\end{equation}
Both operators are nilpotent, $\del_H^2=0=\updelta_H^2$. Let
\begin{equation}
\Delta_H=(\del_H+\updelta_H)^2 \label{DelH}
\end{equation}
be the twisted Laplacian those restriction to $p$-forms will be denoted $\Delta_H^p$. Any $p$-form can be Hodge-decomposed in a sum of twisted exact, twisted coexact and twisted harmonic forms,
\begin{equation}
B_p=B_p^\| + B_p^\bot +\gamma_p, \quad B_p^\|=\del_H B_{p-1}, \quad B_p^\bot=\updelta_H B_{p+1}, \quad \gamma_p \in {\rm ker}\,\Delta_H^p\,.
\end{equation}
The restrictions of $\Delta_H$ to corresponding spaces will be denoted $\Delta_H^{p\|}$ and $\Delta_H^{p\bot}$, respectively. 
$\gamma_p$ denotes twisted harmonic forms. 
Let $\star$ be the normalized Hodge operator, $\star^2=1$. Then, $\star \del_H \star = \updelta_{H^{-1}}$ and
\begin{equation}
\star \Delta_H^p \star = \Delta_{H^{-1}}^{2-p}\,,\qquad \star \Delta_H^{p\|} \star = \Delta_{H^{-1}}^{(2-p)\bot}\,.
\label{dualD}
\end{equation}

Using the formalism above one can rewrite the partition function \eqref{Z2} as a ratio of determinants of twisted Laplacians for 0-forms.
\begin{equation}
Z=
\frac {\det (\Delta_H^0)^{\frac 12}}{
\det (\Delta_{H^{-1}}^0)^{\frac 12}} \label{Z3}
\end{equation}

To evaluate the functional determinants in the partition function \eqref{Z3} we shall use the zeta function definition. Let $D$ be a Laplace type operator on some vector bundle. Then
\begin{equation}
\ln \det D :=-\zeta'_D(1,0)\label{lndet}
\end{equation}
where
\begin{equation}
\zeta_D(h,s):={\rm Tr}\, \big( h D^{-s}\big) \label{defzeta}
\end{equation}
with $h$ being an endomorphism of the bundle (a smooth matrix-valued function). Let us compute the variation
\begin{eqnarray}
\delta_\alpha \zeta_{\Delta_H^0} (1,s)&=&{\rm Tr}\, \big( -s \mat \updelta_H \del_H (\Delta_H^0)^{-s-1}
+s \updelta_H \mat \del_H (\Delta_H^0)^{-s-1}\big)\nonumber\\
&=&-s {\rm Tr}\, \big( \mat (\Delta_H^0)^{-s}\big) + s {\rm Tr}\, \big( \mat (\Delta_H^{1\|})^{-s}\big)\nonumber\\
&=&-s \zeta_{\Delta_H^0}(\mat, s)+s\zeta_{\Delta_H^{1\|}}(\mat, s)\,.
\label{var1}
\end{eqnarray}
Similarly,
\begin{equation}
\delta_\alpha \zeta_{\Delta_{H^{-1}}^0} (1,s)= s \zeta_{\Delta_{H^{-1}}^0}(\mat, s)-s\zeta_{\Delta_{H^{-1}}^{1\|}}(\mat, s)
= s \zeta_{\Delta_{H^{-1}}^0}(\mat, s)-s \zeta_{\Delta_H^{1\bot}}(\mat, s) \label{var2}
\end{equation}
where we have used the duality transformation (\ref{dualD}). The relevant combination of zeta functions varies as
\begin{equation}
\delta_\alpha \big( \zeta_{\Delta_H^0} (1,s)-\zeta_{\Delta_{H^{-1}}^0} (1,s)\big)=
s\big( - \zeta_{\Delta_H^0}(\mat, s) - \zeta_{\Delta_{H^{-1}}^0}(\mat, s) + \zeta_{\Delta_H^{1}}(\mat, s)\big)
\,.\label{var3}
\end{equation}
The right hand side of Eq.\ (\ref{var3}) contains zeta functions of elliptic Laplace type operators. All these zeta functions are regular at $s=0$ and their values at $s=0$ can be expressed through corresponding heat kernel coefficients. For any Laplacian $D$ in two dimensions and any endomorphism  $\mat$,
\begin{equation}
\zeta_D (\mat,0)=a_2(\mat,D) - {\rm Tr}\, \big( \theta \Pi_D \big)\,,\label{zerom}
\end{equation}
where the heat kernel coefficient $a_2$ is calculated below and $\Pi_D$ is the projector on ${\rm ker}\, D$. 

Let us study first the zero mode contribution to the right hand side of (\ref{zerom}). If $D$ is non-negative,
\begin{equation}
\Pi_D=\lim_{t\to \infty} \exp(-tD) \,.
\end{equation}
For $D=\Delta_H^0$ we have
\begin{eqnarray}
{\rm Tr}\ \big( \theta \Pi_{\Delta_H^0} \big)&=&\lim_{t\to\infty} {\rm Tr}\ \big( \theta \exp(-t\Delta_H^0) \big)\nonumber\\
&=&\lim_{t\to\infty} {\rm Tr}\ \big(\Psi H \mat H^{-1} \Psi^{-1} \exp(-tK) \big)={\rm Tr}\ \big(\Psi H \mat H^{-1} \Psi^{-1} \Pi_K \big)\,,
\end{eqnarray}
where $K=\Psi H \Delta_H^0 H^{-1}\Psi^{-1}=\Psi H^2\Psi^{-1}\updelta \Psi H^{-2} \Psi^{-1}\del$. Zero modes of $K$ are constant target space vectors,
so that $\Pi_K$ is just the averaging over $S^2$,
\begin{equation}
{\rm Tr}\ \big( \theta \Pi_{\Delta_H^0} \big)=\frac 1{4\pi}\int d^2x \sqrt{g}\, {\rm tr}\, \big( \Psi H \mat H^{-1} \Psi^{-1}\big)
=\frac 1{4\pi}\int d^2x \sqrt{g}\, {\rm tr}\, \big( \mat \big)=0.\label{Trt}
\end{equation}
Besides, $\dim \ker \Delta_H^0=\dim \ker K=4$. The same conclusions are valid for $\Delta_{H^{-1}}^0$. One may also 
show\footnote{This fact is almost obvious since there are no (untwisted) harmonic one-forms on $S^2$. A formal proof goes like follows.
First, one observes that the combination $\dim \ker \Delta_H^0-\dim \ker \Delta_H^1+\dim\ker\Delta_H^2$ is the index of twisted de Rham complex
\cite{Gilkey:2002qd}, and thus is a homotopy invariant, and thus is equal to $4\chi(S^2)=8$ (where $\chi$ is the Euler characteristic).
As we saw already, $\dim\ker\Delta_H^0=4$ and $\dim\ker\Delta_H^2=\dim\ker\Delta_{H^{-1}}^0=4$. Therefore, $\dim\ker\Delta_H^1=0$. }
that there are no twisted harmonic one forms on $S^2$. We conclude, that there are no zero mode contributions, and
\begin{equation}
\delta_\alpha \big( \zeta'_{\Delta_H^0} (1,0)-\zeta'_{\Delta_{H^{-1}}^0} (1,0)\big)=
-a_2(\mat,\Delta_H^0)-a_2(\mat,\Delta_{H^{-1}}^0) + a_2(\mat,\Delta_H^1) \label{var4}
\end{equation}
The combination of heat kernel coefficients appearing on the right hand side of this equation is called the supertrace of the twisted de Rham complex. It was computed in
\cite{Vassilevich:2000kt} in flat space and in \cite{Gilkey:2002qd} in the case when $H$ is proportional to the unit matrix. 

The heat kernel coefficients may be computed in the following way (see, e.g., \cite{Vassilevich:2003xt} for details).
By a suitable choice of the effective connection $\bar \nabla$ and of the matrix-valued potential $\mathcal{E}$, any operator of Laplace type
can be transformed to the form
\begin{equation}
D=-(\bar \nabla^\mu \bar \nabla_\mu + \mathcal{E})\,.\label{DnabE}
\end{equation}
Then the second heat kernel coefficient on a two-dimensional manifold with curvature $R$ reads
\begin{equation}
a_2(\mat,D)=\frac 1{4\pi} \int \extd^2x \sqrt{g} \,{\rm tr} \left( \mat \big( \mathcal{E} + \tfrac 16 R \big) \right)\,. \label{a2}
\end{equation}
The potential $\mathcal{E}$ may be computed by a standard though lengthy algebra. The shortest way we found is to generalize Lemma 2.2 of Ref.\ \cite{Gilkey:2002hk}. We have for $\Delta_H^0$
\begin{equation}
\mathcal{E}_H^0=\tfrac 12 \hat \nabla^\mu \big( H^{-1} \hat \nabla_\mu H + \hat \nabla_\mu H \cdot H^{-1} \big)
-\tfrac 14 \big( H^{-1} \hat \nabla_\mu H + \hat \nabla_\mu H \cdot H^{-1} \big)^2
-\tfrac 12 \big[ \hat \nabla_\mu H \cdot H^{-1}, H^{-1} \hat \nabla_\mu H \big] \label{EH0}
\end{equation}
For $p=1$, the corresponding Laplacian and matrix-valued potential have vector indices in addition to the target space ones.
\begin{equation}
\mathcal{E}_{H\nu}^{1\ \ \mu}=\delta_\nu^\mu \mathcal{E}_H^0 -R_\nu^{\ \mu}- \hat\nabla^\mu \big( H^{-1} \hat \nabla_\nu H \big) -
\hat\nabla_\nu \big( \hat \nabla^\mu H \cdot H^{-1} \big) + \big[ \hat \nabla_\mu H \cdot H^{-1}, H^{-1} \hat \nabla_\nu H \big]
\label{EH1}
\end{equation}
After collecting everything together we obtain
\begin{equation}
\delta_\alpha \big( \zeta_{\Delta_H^0} (1,s)-\zeta_{\Delta_{H^{-1}}^0} (1,s)\big)=
-\frac 1{4\pi} \int {\rm tr}\, (\mat R)=
-\frac 1{2\pi} \int {\rm tr}\,\mat=0\label{var5}\,,
\end{equation}
which vanishes due to the tracelessness of $\mat$ (\ref{trb}).\footnote{%
The formalism that we have developed above does not allow to compute variations of each of the scalar Laplacians appearing in (\ref{Z3}) separately. The reason is that the restricted operators $\Delta_H^{1\bot}$ and $\Delta_H^{1\|}$ are not Laplace type. Therefore, the corresponding zeta functions are not in general regular at $s=0$, see \cite{Vassilevich:2000kt} for a more detailed discussion.
}

This shows that $\ln Z$ does not depend on $\alpha$. Hence, we get the same result for the one-loop partition function as for $\alpha=0$.
\begin{equation}
Z=1\label{Z1}
\end{equation}
The one-loop partition function is trivial. By an analytic continuation, the same is valid on AdS${}_2$ as well. We have used the first order formulation in the derivation of \eqref{Z1}. To close the potential loophole of quantum inequivalence between first and second order formulations we show in appendix \ref{app:A} that the second order calculation yields the same result.

Using our arguments above on one-loop exactness we have then the result
\eq{
Z_{\rm p} = 1\,.
}{eq:CDV60}
Thus, there are no perturbative corrections to the classical partition function. 

\subsection{Non-perturbative corrections}\label{se:4.3}

Let us finally consider non-perturbative corrections. These come from all classical saddle points consistent with our boundary conditions \eqref{eq:CDV1}, a given periodicity $\beta$ of the boundary coordinate $\varphi$, and smoothness conditions, which we now specify. We allow all smooth Euclidean saddle points; in particular we prohibit conical singularities. Thus, only two saddle points are possible, namely global $\mathbb{H}^2$ \eqref{eq:global} and Poincar{\'e} $\mathbb{H}^2$ \eqref{eq:poincare}. Note that these two saddle points have different topologies: the former is topologically a cylinder, the latter topologically a plane.

Thus, for fixed topology\footnote{%
It is conceivable to sum over both topologies. Then each saddle point contributes with a trivial partition function to the full partition function. However, it would still be a state-independent number and thus of no physical significance.
} there is only one allowed saddle point and we find no instanton corrections.
\eq{
Z_{\rm n} = 1
}{eq:CDV61}

In summary, the results \eqref{eq:CDV59}, \eqref{eq:CDV60} and \eqref{eq:CDV61} together show that the full partition function \eqref{eq:CDV57} is trivial,
\eq{
Z = Z_{\rm c} \times Z_{\rm p} \times Z_{\rm n} = 1\,.
}{eq:CDV62}
We conclude that AdS$_2$ holography is trivial for CDVs not just classically, but also in the full quantum theory, which has only one physical state, the vacuum.

\section{AdS(2) holography for linear dilaton boundary conditions}\label{se:5}

The previous sections dealt with generic models in two dimensions, but very simple vacua --- as we have shown, so simple that they do not allow any physical states, neither classically nor at the quantum level. In this section we consider instead generic linear dilaton vacua, but focus on a very simple model, namely the charged Jackiw--Teitelboim model.

We start by presenting the model and its most important properties in section \ref{se:5.1}. In section \ref{se:5.2} we generalize the linear dilaton boundary conditions of \cite{Grumiller:2013swa}. In section \ref{se:5.3} we calculate the canonical charges, show that they are non-trivial and determine the Virasoro central charge. In section \ref{se:5.4} we discuss thermodynamics and show that the chiral Cardy formula \eqref{eq:CDV28} correctly accounts for the horizon entropy as calculated from Wald's formula or from the Euclidean path integral.

\subsection{Charged Jackiw--Teitelboim model}\label{se:5.1}

We define the model by specifying the free functions in the general action \eqref{eq:CDV19}, \eqref{eq:CDV20} as follows.
\eq{
U=0 \qquad V=-X \qquad F=-1 \qquad \Rightarrow \qquad {\cal V}=X-f^2
}{eq:CDV74}
The gauge theoretic formulation as a PSM as reviewed in section \ref{se:2.1} still applies, of course, but now there is a simpler interpretation as an ordinary (centrally extended) gauge theory, by combining the insights of Isler, Trugenberger, Chamseddine and Wyler \cite{Isler:1989hq, Chamseddine:1989yz} with the ones by Verlinde, Cangemi and Jackiw \cite{Verlinde:1991rf, Cangemi:1992bj}. Namely, consider the connection 1-form $A=e^a P_a+\omega J + a Z$ and define the algebra
\eq{
[P_a,\,P_b] = \epsilon_{ab} J - 2f \epsilon_{ab} Z \qquad [P_a,\, J] = \epsilon_a{}^b P_b 
}{eq:CDV75}
where the central extension generator $Z$ commutes with all other generators. For $f=0$ the $so(2,1)$ formulation of the Jackiw--Teitelboim model \cite{Jackiw:1984, Teitelboim:1984} is recovered \cite{Isler:1989hq, Chamseddine:1989yz}. On the other hand, dropping the term containing the boost generator $J$ in the commutator of two translations yields the centrally extended Poincar\'e algebra \cite{Cangemi:1992bj} that describes the (Weyl rescaled) Witten black hole \cite{Mandal:1991tz, Elitzur:1991cb, Witten:1991yr}, which is a magnetic-like modification of the translation algebra. The charged Jackiw--Teitelboim model \eqref{eq:CDV74} combines both cases and leads to a centrally extended $so(2,1)$ algebra.

For later purposes it is useful to have also the second order form of the action available, including boundary terms. The bulk action can be read off from \eqref{eq:intro0} with the functions as given in \eqref{eq:CDV74}.
\eq{
I_{\textrm{\tiny bulk}} = -\frac{k}{4\pi}\,\int\extd^2x\sqrt{g}\,\big(X(R+2) + \tfrac14\, f^{\mu\nu}f_{\mu\nu}\big)
}{eq:CDV67}
This model has been used by Hartman and Strominger \cite{Hartman:2008dq} and many others (note that we set again the AdS radius to unity, which corresponds to fixing $\ell=2$ in \cite{Hartman:2008dq}, and that we have a different sign in front of the Maxwell term in the action, so that solutions of ${\cal V}=X-f^2=0$ imply positive $X$ for real $f$ in Euclidean signature).

The first proposal for a full action, including boundary terms, can be found in \cite{Castro:2008ms}. Their boundary term depends on the $u(1)$-connection quadratically. However, this boundary action does not generalize to other models. We take instead the general result derived in \cite{Grumiller:2014oha} where the boundary term depends on the field strength non-linearly but on the connection only linearly
\eq{
\Gamma = I_{\textrm{\tiny bulk}} + \frac{k}{4\pi}\,\int\extd^2x\sqrt{g}\,\nabla_\mu\big(f^{\mu\nu}a_\nu\big)  - \frac{k}{2\pi}\,\int\extd x\sqrt{\ga}\,\Big(XK-\sqrt{X^2-\tfrac14\,Xf_{\mu\nu}f^{\mu\nu}}\Big)\,.
}{eq:CDV73}
The action \eqref{eq:CDV73} turns out to have a well-defined variational principle for our boundary conditions specified below, see \eqref{eq:CDV38}. A detailed explanation of the boundary terms in the holographically renormalized action \eqref{eq:CDV73} can be found in \cite{salzer, Grumiller:2014oha}.

[Boundary terms similar to \eqref{eq:CDV73} appear in a large class of models (whenever the electric force is confining at large values of the dilaton), but there are notable families of exceptions \cite{Bagchi:2014ava}. For vanishing or non-confining Maxwell field the remaining boundary terms in \eqref{eq:CDV73} coincide with the ones derived in \cite{Grumiller:2007ju} with the Hamilton--Jacobi method and the ones derived in \cite{Grumiller:2009dx} from local supersymmetry without boundary conditions. It would be interesting to check if the latter derivation works for \eqref{eq:CDV73}. Note that the pre-potential, $u=\sqrt{(X-f^2)^2-f^4}$, is real as long as the dilaton is bounded from below by $X \geq f^2$.]

\subsection{Linear dilaton boundary conditions}\label{se:5.2}

Using again partial gauge fixings and demanding the leading $\vp$-dependent function in the zweibein to be constant we propose the following set of boundary conditions to describe linear dilaton vacua.
\begin{subequations}
 \label{eq:CDV38}
\begin{align}
 X^0 &= X_R(\vp) e^\rho - X_L(\vp)e^{-\rho} & e_{\vp 0} &= \tfrac12\,e^\rho - \tfrac12\, M(\vp) e^{-\rho} & e_{\rho 0} &= 0 \\
 X^1 &= X^1(\vp) & e_{\vp 1} &= 0 & e_{\rho 1} &= 1 \\
 X &= X_R(\vp) e^\rho + \bar f^2 + X_L(\vp) e^{-\rho} & \om_\vp &= -\tfrac12\,e^\rho - \tfrac12\,M(\vp) e^{-\rho} & \om_\rho &= 0 \\
 f &= \bar f & a_\vp &= E\,\om_\vp + j(\vp) & a_\rho &= 0
\end{align}
\end{subequations}
At the moment we allow all free functions and constants appearing in these boundary conditions to be state-dependent.

The boundary conditions \eqref{eq:CDV38} are compatible with the EOM \eqref{eq:CDV7} provided the following relations hold among the free functions. (Some are redundant, but are displayed for later use; $C$ is the non-trivial Casimir of the PSM that corresponds essentially to the mass of a given solution.) 
\begin{subequations}
 \label{eq:CDV41}
\begin{align}
 X^1 &= -2X_R^\prime \\
 X_L &= M X_R + 2 X_R'' \\
 M &=- \frac{X_L^\prime}{X_R^\prime} = \frac{C}{X_R^2} - \frac{2X_R X_R'' - (X_R^\prime)^2}{X_R^2} \label{eq:CDV87} \\
 C &= X_LX_R-\tfrac14\,(X^1)^2 = \textrm{const.}\\
 0 &= X_R M^\prime + 2 X_R^\prime M + 2X_R''' \label{eq:angelinajolie}
\end{align}
\end{subequations}
The relation between mass function $M$ and Casimir $C$ \eqref{eq:CDV87} resembles a twisted Sugawara shift with the dilaton current $\partial_\vp\ln X_R$, but note that there is an additional rescaling with $X_R^2$.

The boundary conditions \eqref{eq:CDV38} are preserved by transformations \eqref{PSMgauge} generated by the following gauge parameters:
\begin{subequations}
 \label{eq:CDV39}
\begin{align}
 \la_0 &= \tfrac12\,\la(\vp)e^\rho - \tfrac12\,\big(M(\vp)\la(\vp)+2\la''(\vp)\big)\, e^{-\rho} \\
 \la_1 &= -\la'(\vp) \\
 \la_X &= -\tfrac12\,\la(\vp)e^\rho - \tfrac12\,\big(M(\vp)\la(\vp)+2\la''(\vp)\big)\, e^{-\rho} \\
 \la_f &= E \la_X + \mu(\vp)
\end{align}
\end{subequations}

The action of the boundary condition preserving transformations \eqref{eq:CDV39} on the various functions appearing in the boundary conditions \eqref{eq:CDV38} yields
\begin{subequations}
 \label{eq:CDV42}
\begin{align}
 \de_{\la,\,\mu} X_R &= X_R \la^\prime - X_R^\prime\la \\
 \de_{\la,\,\mu} X_L &= -M\big( X_R \la^\prime - X_R^\prime\la \big) + 2\big(X_R^\prime\la''-X_R''\la^\prime\big) \\
 \de_{\la,\,\mu} X_1 &= -2\big(X_R \la'' - X_R''\la\big) \\
 \de_{\la,\,\mu} \bar f &= 0 \\
 \de_{\la,\,\mu} M &= -M^\prime\la-2M\la^\prime-2\la''' \label{eq:virasoro} \\
 \de_{\la,\,\mu} j &= -\mu^\prime
\end{align}
\end{subequations}
%An interesting identity that follows from the previous ones is
%\eq{
%X_R \de_\la M + 2 M \de_\la X_R + 2X_R \la''' - 2X_R'''\la = 0\,.
%}{eq:CDV55}
It is encouraging that the mass function $M$ transforms infinitesimally like a chiral component of the stress tensor in a CFT$_2$ \eqref{eq:virasoro}. However, as we saw in the CDV case such a transformation behavior could be pure gauge. In order to see that this is not the case for linear dilaton solutions we consider next the canonical charges.

\subsection{Canonical charges and Virasoro central charge}\label{se:5.3}

We insert the results from the previous subsection into the expression for the canonical currents \eqref{eq:CDV28} evaluated at some cut-off surface $\rho=\rho_c$ and obtain
\eq{
\de Q = \de Q_{\textrm{\tiny int}} + \de Q_{\textrm{\tiny non-int}} +  \de Q_{\textrm{\tiny tot-der}} + \de Q_0 
}{eq:CDV43}
with the individual contributions
\begin{align}
 Q_{\textrm{\tiny int}} &= \frac{k}{2\pi}\,\big(\bar f \mu - M X_R \la - 6 X_R'' \la\big) \\
 \de Q_{\textrm{\tiny non-int}} &= -\frac{k}{2\pi}\,\la M\de X_R \label{eq:CDV76} \\
 Q_{\textrm{\tiny tot-der}} &= -\frac{k}{\pi}\,\partial_\vp\big(X_R \la^\prime-2\lambda X_R^\prime\big) \\
 Q_0 &= {\cal O}(e^{-\rho_c})\,.
\end{align}
Except for one term all contributions to the canonical charges are integrable. An easy way to make the charges integrable is to additionally restrict the boundary conditions \eqref{eq:CDV38} by assuming $X_R=\rm const.$, but then all charges vanish on-shell except for the zero mode charge.\footnote{%
Off-shell the non-zero mode charges are non-trivial even for $X_R=\rm const.$, through the same effect that allowed Kucha\v{r} to make the Schwarzschild mass time-dependent \cite{Kuchar:1994zk}: the Casimir $C$ is then allowed to depend on the boundary coordinate $\vp$. It could be interesting to follow this path, but we will not do so in the present work.
} We proceed below with a much weaker assumption on $X_R$.

In order to deal with the non-integrable part we Fourier transform the essential functions (all sums run over the integers, unless mentioned otherwise) 
\eq{
X_R = \sum_n X_{R\,n} e^{in\vp}\qquad M = \sum_n M_n e^{in\vp}
}{eq:CDV80}
and assume that $M$ is small so that we can expand
\eq{
X_{R\,0} = \bar X + {\cal O}(M^2)\qquad X_{R\,n} = \frac{\bar X}{2n^2}\,M_n + {\cal O}(M^2)\,.
}{eq:CDV81}
The only essential assumption that went into \eqref{eq:CDV81} is that $\bar X$ is a state-independent number; to reduce clutter we assumed that $X_{R\,0}$ has no term linear in $M$. The right equation \eqref{eq:CDV81} is simply a result from solving the EOM to linear order in $M$, see \eqref{eq:angelinajolie}. It implies the relations
\eq{
\partial_\vp X_R = {\cal O}(M) \qquad \delta X_R = {\cal O}(M)\,.
}{eq:CDV86}

Consider now the canonical diffeomorphism current to quadratic order in $M$.
\eq{
\delta Q[\lambda] = -\frac{k}{2\pi}\,\Big(\de (MX_R) + \frac{\bar X}{2}\,\sum_n M_n e^{in\vp} \sum_{m\neq 0} \frac{1}{m^2}\,e^{im\vp}\,\de M_m \Big)\,\la + \partial_\vp(\dots) + {\cal O}(M^3)
}{eq:CDV82}
We are exclusively interested in the zero mode charge, which is why we can neglect total derivative terms even without integrating over $\vp$. The zero mode charge turns out to be integrable to quadratic order in $M$ and reads
\eq{
Q_0[\lambda] = -\frac{k\bar X}{2\pi}\,\Big(M_0 + \frac{3}{2}\,\sum_{n>0} \frac 1{n^2}\, M_{-n} M_n \Big)\,\la + {\cal O}(M^3)\,.
}{eq:CDV83}
The result \eqref{eq:CDV83} can be rewritten as
\eq{
Q_0[\lambda] = -\frac{k\bar X}{2\pi}\,\Big(M - \frac{3}{\bar X^2}\,(\partial_\vp X_R)^2\Big)_0\,\la + {\cal O}(M^3)\,.
}{eq:CDV84}
The first term in \eqref{eq:CDV84} is the expected Virasoro zero-mode and the second term is a Sugawara term for the current $\partial_\vp X_R/\bar X$ (or for $\partial_\vp\ln X_R$, which is the same to this order). In appendix \ref{app:B} we show an exact version of the result \eqref{eq:CDV84} for the zero mode charge in the presence of an ultraviolet cutoff, with essentially the same conclusions.

The leading order term in the zero mode charge \eqref{eq:CDV84} allows to identify the scaling of the Virasoro zero mode.
\eq{
L = \frac{k\bar X}{2\pi}\,M
}{eq:CDV85}
This provides us with a normalization to be used in the Virasoro algebra \eqref{eq:virasoro}.
\eq{
\de L = - L^\prime\la - 2L\la^\prime - \frac{c}{12}\,\la'''
}{eq:CDV48}
Comparing \eqref{eq:CDV48} with \eqref{eq:virasoro} using the relation \eqref{eq:CDV85} establishes the central charge
\eq{
c = 24 k\,\frac{\bar X}{2\pi}\,.
}{eq:CDV49}

\subsection{Entropy, macroscopically and microscopically}\label{se:5.4}

We calculate now entropy in three ways, two macroscopic ones (Euclidean path integral, Wald entropy) and one microscopic one (chiral Cardy formula), and show that all results agree with each other.

We start with the derivation from the Euclidean path integral. To this end we evaluate the full action \eqref{eq:CDV73} on-shell for the boundary conditions \eqref{eq:CDV38} with $X_R=\bar X$ and find
\begin{multline}
\Gamma\big|_{\textrm{\tiny EOM}} = -\frac{k\beta}{4\pi}\,\int\limits_{\rho_0}^{\rho_c}\extd\rho\,e_{\vp 0}\,\big(X(R+2)+\tfrac12\,f_{\rho\vp}^2/e_{\vp 0}^2\big) + \frac{k\beta}{4\pi}\,\int\limits_{\rho_0}^{\rho_c}\extd\rho \,\partial_\rho \big(e_{\vp 0} a_\vp f_{\rho\vp}/e_{\vp 0}^2 \big) \\
- \frac{k\beta}{2\pi}\,\lim_{\rho\to\rho_c}e_{\vp 0}\big(XK - \sqrt{X^2 - \tfrac12\,X f_{\rho\vp}^2/e_{\vp 0}^2}\big)
\label{eq:CDV72}
\end{multline}
where we assumed constant $M$ and the periodicity $\vp\sim\vp+\beta$. 
Using the relations $R=-2$, $X=\bar X e^\rho + \bar f^2 + {\cal O}(e^{-\rho})$, $K = \partial_\rho \ln e_{\vp 0} = 1 + 2M e^{-2\rho} + {\cal O}(e^{-4\rho})$, $M = 4\pi^2/\beta^2$, $e^{\rho_0}=2\pi/\beta$, $f_{\rho\vp}=-E e_{\vp 0}$ and taking the limit $\rho_c\to\infty$ then establishes a finite result for the on-shell action
\eq{
\Gamma\big|_{\textrm{\tiny EOM}} = -\frac{k}{4}\, E^2 - 2\pi k \bar X T \,.
}{eq:CDV68}
The free energy is the on-shell action \eqref{eq:CDV68} times temperature $T$.
\eq{
F(E,\,T) = -\frac{k}{4}\, E^2 T -2\pi k \bar X T^2
}{eq:CDV69}
Entropy is derived in the usual way, $S=-\partial F/\partial T|_E$, yielding
\eq{
S = 4\pi k \bar X T + \frac{k}{4}\,E^2 \,.
}{eq:CDV70}

Wald's method leads to the result that entropy is essentially given by the dilaton evaluated at the horizon \cite{Gegenberg:1994pv}
\eq{
S_{\textrm{\tiny Wald}} =  k X_h\,.
}{eq:CDV79}
The locus of the horizon $\rho=\rho_h$ is determined by the zero of $e_{\vp 0}$, which yields $e^{2\rho_h}=M$. We then obtain
\eq{
X_h = 2\bar X\sqrt{M} + \bar f^2\,.
}{eq:CDV52}
With the relations $M=4\pi^2T^2$ and $\bar f^2=E^2/4$ we see that the macroscopic entropies \eqref{eq:CDV70} and \eqref{eq:CDV79} coincide with each other.

Let us now check if we can trust the result for the Virasoro central charge \eqref{eq:CDV49} by calculating the entropy using the Cardy formula \eqref{eq:CDV27} and verifying whether it agrees with the macroscopic result \eqref{eq:CDV70} for vanishing $E$ . We find
\eq{
S_{\textrm{\tiny Cardy}} = \frac{\pi^2 c\, T}{3} = 2\pi\sqrt{\frac{c L}{6}} = 2k\bar X \sqrt{M} = 4\pi k\bar XT\,.
}{eq:CDV50}
The microscopic result \eqref{eq:CDV50} coincides precisely with the macroscopic result \eqref{eq:CDV70} for $E=0$. This observation was already made in \cite{Grumiller:2013swa}, but without resolving the non-integrability of the canonical zero-mode charge.

\section{Discussion}\label{se:6}

We have shown the triviality of AdS$_2$ holography for constant dilaton (see section \ref{se:2}) in two ways, by calculating the canonical charges in section \ref{se:3} and by performing the quantum gravity path integral in section \ref{se:4} (see also appendix \ref{app:A}). These triviality results concur with earlier observations, see e.g.~\cite{Maldacena:1998uz, Balasubramanian:2009bg, Castro:2014ima}, but apply now to all models \eqref{eq:intro0} and also to the various generalizations we discussed in section \ref{se:3.3}, including non-linear interactions with the Maxwell field, higher-spin generalizations and generalizations to Yang--Mills.

For a specific model, the charged Jackiw--Teitelboim model \eqref{eq:CDV67}, we considered linear dilaton boundary conditions in section \ref{se:5} and found that they lead to non-trivial canonical currents. In general they are not obviously integrable; however, the zero mode charge turned out to be integrable in a perturbative expansion up to quadratic order in the mass, and integrable non-perturbatively in the presence of an ultraviolet cutoff, see appendix \ref{app:B}. 
It would be interesting to check under which conditions (if any) all diffeomorphism charges are integrable to all orders in the mass. Moreover, it could be rewarding to calculate the one-loop partition function of the charged Jackiw--Teitelboim model to verify if it coincides with the chiral Virasoro character. 

We showed that our result for the Virasoro central charge \eqref{eq:CDV49} yields the correct entropy \eqref{eq:CDV50} using the chiral Cardy formula and comparing it with results from the Euclidean path integral or the Wald entropy. Thus, we have an explicit example of non-trivial AdS$_2$ holography, provided we consider a non-constant dilaton.

Most likely the non-triviality of the linear dilaton sector extends to fairly generic dilaton gravity models \eqref{eq:intro0} as long as they allow AdS$_2$ holography and possibly even for more general holographic setups. It would be interesting to verify this.

\section*{Acknowledgments}

JS thanks Stefan Prohazka and Friedrich Sch\"oller for discussions. 

DG was supported by the START project Y 435-N16 of the Austrian Science
Fund (FWF), the FWF projects I 952-N16, I 1030-N27 and P 27182-N27, and by the program Science without Borders, project CNPq-401180/2014-0. 

JS was supported by the FWF projects Y 435-N16 and P 27182-N27.

DV was supported by the FAPESP project 2012/00333-7, by the CNPq project 306208/2013-0, and by the program Science without Borders, project CNPq 401180/2014-0. 

DG and DV thank each others home institutions for hosting extended research visits during this project.

\begin{appendix}

\section{One-loop result in second order formulation}\label{app:A}

In order to close possible loopholes in our argument due to potential quantum inequivalence of first order formulation \eqref{eq:CDV19} and second order formulation \eqref{eq:intro0}, we demonstrate in this appendix the triviality of constant dilaton holography on $\mathbb{H}^2$ using the latter. 

We split again the partition function $Z$ into classical contributions (c), perturbative contributions (p) and non-perturbative contributions (n):

\begin{equation}
  \label{eq:22}
  Z=Z_{\rm c} \times Z_{\textrm p} \times Z_{\textrm n}\, .
\end{equation}
Using the bulk action \eqref{eq:intro0} supplemented by the boundary term presented in \cite{Grumiller:2014oha} such that the action yields a well-defined variational principle, the classical contribution $Z_{\textrm c}$ is evaluated to be a state-independent constant. Thus, after a renormalization of the action we can set
\begin{equation}
  \label{eq:23}
  Z_{\textrm c}=1\, .
\end{equation}
Based on the same arguments brought forth in the main text we conclude that
\begin{equation}
  \label{eq:24}
  Z_{\textrm n}=1\, ,
\end{equation}
as well. Again the only potential contribution comes from the perturbative corrections $Z_{\textrm p}$ that we calculate in the following. 

However, due to the same reasons mentioned in the main text after equation \eqref{expan}, we find it more convenient to calculate the partition function on $S^2$ and analytically continue the result to $\mathbb{H}^2$.

The first variation of \eqref{eq:intro0} yields the following EOM:
\begin{subequations}
\begin{equation}
\label{eq:Aeom1}
R+U'(\nabla X)^2+2U\nabla^2X-2V'-\frac{1}{4}F'f_{\mu\nu}f^{\mu\nu}=0
\end{equation}
\begin{equation}
\label{eq:Aeom2}
\nabla^\mu\nabla^\nu X-g^{\mu\nu}\nabla^2X+U(\nabla^\mu\nabla^\nu X-\frac{1}{2}g^{\mu\nu}(\nabla X)^2)
-g^{\mu\nu}(V-\frac{1}{8}Ff^2)=0
\end{equation}
\begin{equation}
\label{eq:Aeom3}
\nabla_{\nu}(Ff^{\mu\nu})=0
\end{equation}
\end{subequations}
 From the last equation one sees again that $f_{\mu\nu}$ has the form
 \begin{equation}
   \label{eq:1}
   f_{\mu\nu}=-E \epsilon_{\mu\nu}\, ,
 \end{equation}
where $\epsilon_{\mu\nu}$ corresponds to the Levi-Civit\'a tensor, in concurrence with equation \eqref{eq:E}. Since the dilaton field $X$ is constant
\begin{equation}
  \label{eq:2}
  X=\bar{X}
\end{equation}
one obtains again the following restrictions on the the functions $V$ and $F$
\begin{subequations}
  \begin{equation}
    \label{eq:3}
    V(\bar{X})=\frac{1}{4}F(\bar{X})E^2
  \end{equation}
\begin{equation}
\label{eq:3a}
V'(\bar{X})+\frac{1}{4}F'(\bar{X})E^2=1
\end{equation}
\end{subequations}
These equations are equivalent to the conditions \eqref{eq:CDV4} and \eqref{eq:CDV5}, but modified to account for the fact that we consider now $S^2$. 

The second variation of \eqref{eq:intro0} is straightforward and produces the following terms after having made use of the on-shell relations \eqref{eq:Aeom1}-\eqref{eq:Aeom3}
\begin{subequations}
\label{eq:varGamma}
\begin{align}
\delta^{(2)}I=&
\begin{aligned}
\label{XX}
-\frac{k}{4\pi}\int\extd^2x\sqrt{\bar{g}}\, X\left(2\bar{U}\nabla^2-2\bar{V}''-\frac{1}{2}\bar{F}''E^2\right) X
\end{aligned}
\\
&
\begin{aligned}
\label{AA}
-\frac{k}{4\pi}\int\extd^2x\sqrt{\bar{g}}\, A_\mu\bar{F}\left(\bar{g}^{\mu\nu}\nabla^2-\nabla^\nu\nabla^\mu\right) A_\nu
\end{aligned}
\\
&
\begin{aligned}
\label{gg}
-\frac{k}{4\pi}\int\extd^2x\sqrt{\bar{g}}\, g_{\mu\nu}\left(-\frac{1}{4}\bar{g}^{\mu\nu}\bar{g}^{\alpha\beta}\bar{F} E^2\right) g_{\alpha\beta}
\end{aligned}
\\
&
\begin{aligned}
\label{gX}
-\frac{k}{2\pi}\int\extd^2x\sqrt{\bar{g}}\, g_{\mu\nu}\left(\nabla^{\mu}\nabla^{\nu}-\bar{g}^{\mu\nu}\nabla^2-\bar{g}^{\mu\nu}\bar{V}'+\frac{1}{4}\bar{g}^{\mu\nu}\bar{F}'E^2\right) X
\end{aligned}
\\
&
\begin{aligned}
\label{gA}
-\frac{k}{2\pi}\int\extd^2x\sqrt{\bar{g}}\, g_{\mu\nu}\left(-\bar{g}^{\mu\nu}\frac{1}{2}\bar{F}E \epsilon^{\alpha\beta}\nabla_\alpha\right)\, A_\beta
\end{aligned}
\\
&
\begin{aligned}
\label{XA}
-\frac{k}{2\pi}\int\extd^2x \sqrt{\bar{g}} \, X\left(-\bar{F}'E\epsilon^{\mu\nu}\nabla_\nu\right)\, A_\mu
\end{aligned}
\end{align}
\end{subequations}
Here and in the following barred quantities denote background fields that solve the EOM and obey the conditions \eqref{eq:2}-\eqref{eq:3a} while unbarred quantities denote fluctuations. 

Consider first the case without $U(1)$ field, i.e. $F=0$. Then the only contributions come from \eqref{XX} and \eqref{gX}. It is convenient to decompose the metric fluctuations as
\begin{equation}
  \label{eq:4}
   g_{\mu\nu}=\frac{1}{2}\bar{g}_{\mu\nu}h+\nabla_{\mu}\xi_\nu+\nabla_\nu \xi_\mu\, ,
\end{equation}
so that the second variation simplifies to 
\begin{subequations}
\begin{align}
\delta^{(2)}I=&
\begin{aligned}
\label{XX2}
-\frac{k}{4\pi}\int\extd^2x\sqrt{g}\,\, X\left(2\bar{U}\nabla^2-2\bar{V}'')\right) X
\end{aligned}
\\
&
\begin{aligned}
\label{gX2}
+\frac{k}{4\pi}\int\extd^2\sqrt{g}\, \, h\left(\nabla^2+2\right) X
\end{aligned}
\end{align}
\end{subequations}
Notice that the dependence on the diffeomorphisms $\xi_\mu$ drops out. 

We are interested in evaluating the path integral
\begin{equation}
  \label{eq:Zpdef}
  Z_{\textrm p}=V^{-1}_{\textrm {\tiny gauge}}\int \mathcal{D} X \mathcal{D}g \exp{(-\delta^{(2)}I)}\, , 
\end{equation}
where $V^{-1}_{\textrm {\tiny gauge}}$ corresponds to the (infinite) volume of the gauge group.

The path integral measure is defined again by a condition equivalent to \eqref{pint}, where we take the inner product of scalars $X$, vectors $\xi_\mu$ and tensors $g_{\mu\nu}$, respectively, to be the ultralocal products
\begin{align}
  \label{eq:5}
  \langle X,X'\rangle&=\int\extd^2x\sqrt{\bar{g}}\,X\,X'\, ,
 \\
 \label{eq:6}
  \langle\xi,\xi'\rangle&=\int\extd^2x\sqrt{\bar{g}}\,\bar{g}^{\mu\nu}\,\xi_\mu\,\xi'_{\nu}\, ,
\\
\label{eq:7}
\langle g, g \rangle&=\int\extd^2x\sqrt{\bar{g}}\,\bar{g}^{\mu\alpha}\,\bar{g}^{\nu\beta}\,g_{\mu\nu}\,g_{\alpha \beta}\, .
\end{align}
The change in variables \eqref{eq:4} yields a Jacobian factor $Z_{\rm gh}$ that corresponds to a (Faddeev-Popov) ghost determinant (cf. \cite{Vassilevich:2003xt,Gaberdiel:2010xv} for further details) 
\begin{equation}
  \label{eq:8}
  \mathcal{D}g_{\mu\nu}=Z_{\rm gh}\mathcal{D}h\mathcal{D}\xi_\mu\, .
\end{equation}
This determinant is most easily calculated in two steps. First, $\xi_\mu$ can be further decomposed in exact and co-exact contributions
\begin{equation}
\label{eq:xideco}
 \xi_{\mu}=\nabla_{\mu}\sigma_1+\epsilon_{\mu}{}^{\lambda}\nabla_{\lambda}\sigma_2\, .
 \end{equation}
Notice that the harmonic contribution is absent as there are no harmonic one-forms on $S^2$. This simplification is the reason we evaluate the partition function on $S^2$. 

Using the definition of the measure with \eqref{eq:6}
we find that the transformation $\mathcal{D}\xi=J_1\mathcal{D}\sigma_1\mathcal{D}\sigma_2$ yields the Jacobian
\begin{equation}
  \label{eq:J_1}
  J_1=\det{(\nabla^2)}^{1/2}_0\det{(\nabla^2)}^{1/2}_0\, ,
\end{equation}
 where the subscript $0$ denotes determinants over scalar fluctuations. 

The metric fluctuation is now
\begin{equation}
  \label{eq:9}
  h_{\mu\nu}=\frac{1}{2}g_{\mu\nu}h-g_{\mu\nu}\nabla^2\sigma_1 +2\nabla_{\mu}\nabla_{\nu}\sigma_1+\epsilon_{\nu}{}^{\lambda}\nabla_{\lambda}\nabla_{\mu}\sigma_2+\epsilon_{\mu}{}^{\lambda}\nabla_{\lambda}\nabla_{\nu}\sigma_2\, 
\end{equation}
where we shifted the trace $h\rightarrow h-2\nabla^2\sigma_1$ which produces a unit Jacobian factor.

By using again the definition of the measure, we see that the  decomposition \eqref{eq:9} induces the change $\mathcal{D}h_{\mu\nu}=J_2\mathcal{D}h\mathcal{D}\sigma_1\mathcal{D}\sigma_2$ with 
\begin{equation}
  \label{eq:10}
 J_2=\det{(\nabla^2)}^{1/2}_0\det{(\nabla^2)}^{1/2}_0\det{(\nabla^2+2)}_0\, ,
\end{equation}
where the $2$ in the last determinant comes from the Ricci scalar of the sphere. 
Thus, we find the ghost determinant $Z_{\rm gh}$ to be
\begin{equation}
  \label{eq:11}
  Z_{\rm gh}=J_2/J_1=\det(\nabla^2+2)_0\, .
\end{equation}
The path integral \eqref{eq:Zpdef} can therefore be written as
\begin{equation}
  \label{eq:12}
  Z_{\textrm p}=V^{-1}_{\textrm {\tiny gauge}}Z_{\rm gh}\int\mathcal{D}X\mathcal{D}h\mathcal{D}\mathcal{\xi}\exp{(-\delta^{(2)}I)}\,. 
\end{equation}
The integration over $\xi_\mu$ can be performed trivially and cancels $V^{-1}_{\textrm {\tiny gauge}}$. Due to the vanishing of the $h,h$ variation in $\delta^{(2)}I$ only the two off-diagonal terms \eqref{gX2} contribute and the perturbative contribution to the partition function is given by
\begin{equation}
  \label{eq:13}
  Z_{\rm p}=Z_{\rm gh}\det{(\nabla^2+2)}^{-1}_0=\frac{\det{(\nabla^2+2)}_0}{\det{(\nabla^2+2)}_0}=1\,.
\end{equation}
 
Thus, for vanishing electric field the partition function is trivial on $S^2$. By analytic continuation the same is valid on $\mathbb{H}^2$. 

In the case of non-vanishing $U(1)$ field it is convenient to decompose the field $A_\mu$ in a way similar to \eqref{eq:xideco}
\begin{equation}
  \label{eq:14}
  A_{\mu}=\nabla_\mu\alpha_1+\epsilon_{\mu}^{\,\, \lambda}\nabla_\lambda\alpha_2\, .
\end{equation}
The second variation \eqref{eq:varGamma} thus reads
\begin{subequations}
\begin{align}
  \delta^{(2)}I=&
\begin{aligned}
-\frac{k}{4\pi}\int\extd^2x\sqrt{\bar{g}}\, X\left(2\bar{U}\nabla^2-2\bar{V}''-\frac{1}{2}\bar{F}''E^2\right)\,X
\end{aligned}
\\
&
\begin{aligned}
-\frac{k}{4\pi}\int\extd^2x\sqrt{\bar{g}}\,\tilde{\alpha}_2(-\bar{F})\tilde{\alpha}_2
\end{aligned}
\\
&
\begin{aligned}
-\frac{k}{4\pi}\int\extd^2x\sqrt{\bar{g}}\,\tilde{h}(-\frac{1}{4}\bar{F}E^2)\tilde{h}
\end{aligned}
\\
&
\begin{aligned}
-\frac{k}{2\pi}\int\extd^2x\sqrt{\bar{g}}\,\tilde{h}(-\frac{1}{2}\nabla^2-\bar{V}'+\frac{1}{4}\bar{F}'E^2) X
\end{aligned}
\\
&
\begin{aligned}
-\frac{k}{2\pi}\int\extd^2x\sqrt{\bar{g}}\,\tilde{h}(\frac{1}{2}\bar{F}E)\tilde{\alpha}_2
\end{aligned}
\\
&
\begin{aligned}
-\frac{k}{2\pi}\int\extd^2x\sqrt{\bar{g}}\, X(-\bar{F}'E)\tilde{\alpha}_2
\end{aligned}
\end{align}
\end{subequations}
Here we introduced the shifted trace $\tilde{h}=h-2\nabla^2\sigma_1$ and set $\tilde{\alpha}_2=\nabla^2\alpha_2$. It is evident that the second variation is independent of $\xi_\mu$ and the longitudinal part $\alpha_1$ of $A_\mu$. Furthermore, the change of variables
\begin{subequations}
\label{eq:psichi}
\begin{align}
  \psi&=\frac{2E}{4+E^2}\left(\frac{2}{E}\tilde{h}+\tilde{\alpha}_2\right)
\\
  \chi&=\frac{4-E^2}{4+E^2}\left(-\frac{E}{2}\tilde{h}+\tilde{\alpha}_2\right)
\end{align}
\end{subequations}
brings the second variation of $I$ into the convenient form
\begin{subequations}
\label{eq:convi}
\begin{align}
  \delta^{(2)}I=&
\begin{aligned}
-\frac{k}{4\pi}\int\extd^2x\sqrt{\bar{g}}\, X\left(2\bar{U}\nabla^2-2\bar{V}''-\frac{1}{2}\bar{F}''E^2\right) X
\end{aligned}
\\
&
\begin{aligned}
-\frac{k}{4\pi}\int\extd^2x\sqrt{\bar{g}}\, \chi\left(-\bar{F}\right)\chi
\end{aligned}
\\
&
\begin{aligned}
\label{eq:relevant}
+\frac{k}{4\pi}\int\extd^2x\sqrt{\bar{g}}\,\, \psi\left(\nabla^2+2\right) X
\end{aligned}
\\
&
\begin{aligned}
-\frac{k}{2\pi}\int\extd^2x\sqrt{\bar{g}}\,\chi\left(\frac{4-E^2}{2E}\right)\left(\frac{1}{2}\nabla^2+\bar{V}'-\frac{1}{4}\bar{F}'(E^2-8)\right) X\, .
\end{aligned}
\end{align}
\end{subequations}

The perturbative contribution to the partition function for non-vanishing $U(1)$ field is given by
\begin{equation}
  \label{eq:15}
  Z_{\textrm p}=V^{-1}_{\textrm {\tiny gauge}}\int\mathcal{D}X\mathcal{D}g\mathcal{D}A \exp{(-\delta^{(2)}I)}\, .
\end{equation}
However, in order to evaluate it one has to determine the Jacobian for the transformation of variables $(g_{\mu\nu},A_{\mu},X) \rightarrow (\psi,\chi,\xi,\alpha_1,X)$
\begin{equation}
  \label{eq:17}
  \mathcal{D}g\mathcal{D}A\mathcal{D}X=Z_{\rm gh,A}\mathcal{D}\psi\mathcal{D}\chi\mathcal{D}\xi\mathcal{D}\alpha_1\mathcal{D}X
\end{equation}
 The decomposition \eqref{eq:14} induces the change of variables
\begin{equation}
  \label{eq:16}
  \mathcal{D}A_\mu=J_3\mathcal{D}\alpha_1\mathcal{D}\alpha_2\, ,
\end{equation}
where $J_3$ is given by the same expression as $J_1$ \eqref{eq:8}. Thus, one finds
\begin{equation}
  \label{eq:18}
  \mathcal{D}g\mathcal{D}A\mathcal{D}X=Z_{\rm gh}J_3\mathcal{D}h\mathcal{D}\xi\mathcal{D}\alpha_1\mathcal{D}\alpha_2\mathcal{D}X=Z_{\rm gh}J_3(\det{(\nabla^2)})_0^{-1}\mathcal{D}h\mathcal{D}\xi\mathcal{D}\alpha_1\mathcal{D}\tilde{\alpha}_2\mathcal{D}X\, .
\end{equation}
Taking into account that the redefinition $h \rightarrow \tilde{h}$ and the transformation \eqref{eq:psichi} both yield a unit Jacobian, the ghost determinant $Z_{\rm gh,A}$ for the change of variables \eqref{eq:17} is given by
\begin{equation}
  \label{eq:19}
  Z_{\rm gh,A}=\frac{Z_{\rm gh}J_3}{\det{(\nabla^2)_0}}\,. 
\end{equation}

Consequently, the path integral \eqref{eq:15} is
\begin{equation}
  \label{eq:20}
  Z_{\textrm p}=V^{-1}_{\textrm {\tiny gauge}}\frac{Z_{\textrm {\tiny gh}}J_3}{\det(\nabla^2)_0}\int\mathcal{D}\psi\mathcal{D}\chi\mathcal{D}X\mathcal{D}\xi\mathcal{D}\alpha_2\exp{(-\delta^{(2)}I)}\,. 
\end{equation}
Due to the form of the matrix of fluctuations \eqref{eq:convi} the only non-vanishing contribution comes from the $\psi,X$ off-diagonal terms \eqref{eq:relevant}, while the other integrations being trivial cancel the gauge volume. The final result is thus given by
\begin{equation}
  \label{eq:21}
  Z_{\textrm p}=\frac{\det(\nabla^2+2)_0\det(\nabla^2)_0^{1/2}\det(\nabla^2)_0^{1/2}}{\det(\nabla^2+2)_0\det(\nabla^2)_0}=1\, .
\end{equation}
This is the result for $S^2$ but, again, by analytic continuation it is true for $\mathbb{H}^2$, as well. 

Thus, we recover the result \eqref{eq:CDV60}, obtained in the main text using the first order formulation, that there are no perturbative corrections to the partition function. 
%\newpage

\section{Exact result for zero mode charge}\label{app:B}

In this appendix we derive an exact expression for the (potentially) non-integrable contribution to the zero mode charge \eqref{eq:CDV76} that essentially coincides with the perturbative result in the main text.

We start here with the Fourier decomposition \eqref{eq:CDV80} and the on-shell relation \eqref{eq:angelinajolie}. Together they imply an infinite set of linear equations for the Fourier coefficients $X_{R\,n}$ in terms of $M_n$. To simplify the notation we rename $X_{R\,n}$ as $X_n$.
\eq{
n^3 X_n = \frac12\,\sum_m (n+m) X_m M_{n-m}
}{eq:CDV88}

We solve now this system of equations assuming that there is an ultraviolet cutoff, in the sense that the Fourier components of the mass function vanish when the absolute value of the index is sufficiently big.
\eq{
M_n = 0 \quad \forall \,|n| > N
}{eq:CDV89}
Our derivation does not depend on the value of $N$, but it does require some cutoff. We assume that we have a tight bound on $N$,
\eq{
M_N \neq 0 \neq M_{-N}\,.
}{eq:CDV90}
One of these inequalities can always achieved with no loss of generality; the second one is assumed for simplicity and could be generalized.

The (potentially) non-integrable zero-mode contribution to the canonical charge \eqref{eq:CDV76} is then given by
\eq{
\delta Q_{\textrm{\tiny non-int}\,0} = -\frac{k}{2\pi}\,\la\,\big(M_{-N}\delta X_N + M_{-N+1}\delta X_{N-1}+ \dots + M_N\delta X_{-N}\big)\,.
}{eq:CDV91}
Therefore, we need to know only the coefficients $X_n$ with $|n|\leq N$. 

Inspired by the perturbative results in the main text let us use the Ansatz
\eq{
X_0 = \bar X\qquad X_n = \frac{\bar X}{2n^2}\,M_n\quad\textrm{if}\; N\geq |n|\neq 0\,.
}{eq:CDV92}
As opposed to the main text the relations \eqref{eq:CDV92} are now exact.
The condition \eqref{eq:CDV88} then holds for $n=0$.
\eq{
0 = \frac12\,\sum_{m=1}^N m X_m M_{-m} + \frac12\,\sum_{m=-1}^{-N} m X_m M_{-m}
}{eq:CDV93}
We prove now that the ansatz \eqref{eq:CDV92} is compatible with the full set of equations \eqref{eq:CDV88}, provided we fix the remaining Fourier coefficients $X_n$ appropriately for $|n|>N$.

For $n=1$ \eqref{eq:CDV88} yields a linear equation for the Fourier coefficient $X_{N+1}$:
\eq{
X_1 = \frac12\,\big(M_{-N} X_{N+1} (N+2) + M_{-N+1} X_N (N+1) + \dots + M_N X_{-N+1} (-N+2)\big)
}{eq:CDV94}
Note that all terms in this equation are known, except for the first one on the right hand side, which then determines $X_{N+1}$. Similarly, for increasing values of $n$ we can iteratively determine $X_{N+n}$ in this way. The calculations for negative $n$ are completely analogous and allow to determine iteratively the coefficients $X_{-N-|n|}$. Thus, all the equations \eqref{eq:CDV88} are solved exactly in this simple way.

The coefficients $X_n$ with $|n|>N$ do not have the form \eqref{eq:CDV92}, thereby differing from our perturbative result \eqref{eq:CDV81}. However, none of them contributes to the canonical zero mode charges anyhow.
\eq{
Q_{\textrm{\tiny non-int}\,0} =  -\frac{k\bar X}{4\pi}\,\la\,\sum_{n=1}^N \frac{1}{n^2}\,M_{-n} M_n = -\frac{k\bar X}{4\pi}\,\la\,\sum_{n>0} \frac{1}{n^2}\,M_{-n} M_n
}{eq:CDV95}

Including the integrable contributions the full result for the zero mode diffeomorphism charge reads
\eq{
Q_0[\lambda] = -\frac{k\bar X}{2\pi}\,\Big(M_0 + \frac{3}{2}\,\sum_{n>0} \frac 1{n^2}\, M_{-n} M_n \Big)\,\la \,.
}{eq:CDV96}
The final result \eqref{eq:CDV96} for the exact zero mode charge is now true to all orders in the mass $M$ and is essentially equivalent to the perturbative result \eqref{eq:CDV83}.

\end{appendix}

\addcontentsline{toc}{section}{References}

% \bibliographystyle{fullsort}
% \bibliography{dv,review}

\end{document}